\documentclass{ws-ijmpe}

\begin{document}

\markboth{Francesca Sammarruca}{The Microscopic Approach to Nuclear Matter 
and Neutron Star Matter}

\catchline{}{}{}{}{}

\title{THE MICROSCOPIC APPROACH TO NUCLEAR MATTER \\
AND NEUTRON STAR MATTER                              
}

\author{\footnotesize FRANCESCA SAMMARRUCA                              
}

\address{Physics Department, University of Idaho\\
Moscow, Idaho 83844, U.S.A. \\ 
fsammarr@uidaho.edu}

\maketitle

\begin{history}
\received{(received date)}
\revised{(revised date)}
\end{history}

\begin{abstract}

We review a variety of theoretical and experimental investigations 
aimed at improving our knowledge of the 
nuclear matter equation of state. Of particular interest are nuclear matter extreme states            
in terms of density and/or isospin asymmetry. 
The equation of state of matter with unequal concentrations of protons and 
neutrons 
has numerous applications. These include heavy-ion collisions,                  
the physics 
of rare, short-lived nuclei and, on a dramatically different scale,
the physics of neutron stars. The ``common denominator" among these (seemingly) very different systems is the symmetry energy, 
which plays a crucial role in both the formation of the neutron skin in neutron-rich
nuclei and the radius of a neutron star (a system 18 orders of magnitude
larger and 55 orders of magnitudes heavier). 
The details of the density dependence of the symmetry energy are not yet sufficiently 
constrained.         
Throughout this article, 
our emphasis will be on the importance of adopting a microscopic approach to the many-body problem, which we believe to be the one with true predictive power. 

\end{abstract}

\newpage
\section{Introduction}

The interaction of hadrons in nuclei is a problem that goes to the very core of 
nuclear physics. In fact, our present knowledge of the nuclear force in free space is, in itself, 
the result of decades of struggle,\cite{Mac89} which will not be reviewed in this article.  
The nature of the nuclear force in the medium is of course a much more complex problem, 
as it involves aspects of the force that cannot be constrained
through free-space nucleon-nucleon (NN) scattering. Predictions of properties of nuclei are 
the ultimate test for many-body theories. 

Nuclear matter is an alternative and convenient theoretical laboratory for many-body theories. By ``nuclear matter" we mean an infinite system 
of nucleons acted on by their mutual strong forces and no electromagnetic interactions. Nuclear matter 
is characterized by its energy/particle as a function of density and other thermodynamic quantities, if 
appropriate (e.g.~temperature). Such relation is known as the nuclear matter equation of state (EoS). 
The translational invariance of the system facilitates theoretical calculations. At the same time, adopting     
what is known as 
``local density approximation", one can use the EoS directly in calculations of finite systems. This procedure 
is applied,   
for instance, in Thomas-Fermi calculations within the liquid drop model, where an appropriate energy functional is 
written in terms of the EoS.\cite{Oya98,Furn,SL09} 

 Symmetric nuclear matter (that is, matter with equal densities of protons and
neutrons) has been studied extensively. The so-called conventional
 approach goes back to earlier works by Brueckner and others\cite{5,6,7,8} and is known 
 as the Brueckner-Hartree-Fock (BHF) theory.                                                     
The Brueckner theory is based on a linked-cluster perturbation series of the ground state energy 
of a many-body system.\cite{5,6,7,8,HT70,10} Such series was shown to converge when the cluster diagrams are 
regrouped according to the number of hole lines. 
The variational approach was also pursued as an 
alternative method\cite{11,12} and yielded predictions in close agreement with those from Brueckner theory         
when realistic NN potentials were employed.\cite{13} 

 During the 1980's, the Dirac-Brueckner-Hartree-Fock (DBHF) approach
was developed.\cite{14,15,BM84} The main 
 break-through came with the observation that the DBHF theory, unlike       
 conventional Brueckner theory, 
 could describe successfully the saturation properties of nuclear matter, that
 is, saturation energy and density of the equation of state.             
The DBHF method adopts realistic NN interactions and contains important relativistic features.
It describes the nuclear mean field in terms of 
strong, competing scalar and vector fields that, together, account for the binding
of nucleons as well as the large spin-orbit splitting seen in nuclear states.

Regardless of the chosen many-body theory, a quantitative NN potential should be part of its input. 
Recently, chiral effective theories of the nuclear force\cite{chi} have become popular as a mean to respect 
the symmetries of Quantum Chromo Dynamics (QCD) while retaining the basic degrees of freedom typical of low-energy nuclear physics. 
Chiral effective theories provide a well-defined scheme to determine the appropriate many-body 
diagrams to be included at each order of the perturbation. However, being based on a low-momentum 
expansion, interactions derived from chiral perturbation theory are not suitable for applications in dense nuclear/neutron
matter, where high Fermi momenta are involved. Instead,                                      
meson-theoretic or phenomenological NN potentials are typically employed 
as input to the many-body theory.                      
Mean-field models (relativistic and non-relativistic) 
are a popular, although non-microscopic, alternative to methods based on the in-medium reaction matrix, such 
as BHF and DBHF. 

It is the purpose of this article to review the status of microscopic studies of nuclear and 
neutron-rich matter, with particular emphasis on the latter, as it relates 
to present empirical investigations within terrestrial laboratories 
(heavy-ion collisions) or astrophysical observations (neutron stars). At the same time, we will review the 
present status concerning available empirical information which can be utilized to guide and constrain 
theories. 
We also wish to provide a self-contained account of the recent work with asymmetric matter done by the Idaho group,
including extensive numerical tables for the interested user.  

Clearly, the goal                         
to describe the properties of (dense) many-body systems 
consistently from the underlying forces {\it and} including all potentially important mechanisms, 
is a  most ambitious program and far from having been completed. The importance of pursuing a microscopic approach towards the accomplishment of 
this goal is a theme      
that will surface repeatedly throughout this article.                                     

\section{Asymmetric Nuclear Matter: Some Facts and Phenomenology}

Isospin-asymmetric nuclear matter (IANM) simulates the interior of a ``nucleus" with unequal densities of protons and neutrons.
The equation of state of (cold) IANM is then a function of density as well as the relative concentrations 
of protons and neutrons. 

The recent and fast-growing interest in IANM stems from its close connection to neutron-rich nuclei, or,
more generally, asymmetric nuclei, including the very ``exotic" ones known as ``halo" nuclei. 
At this time, the boundaries of the nuclear chart are uncertain, with a few hundreds stable nuclides 
known to exist and perhaps a few thousands believed to exist. 
The Facility for Rare Isotope Beams (FRIB) has recently been approved for design and construction at 
Michigan State University (MSU). 
The facility will deliver intense beams of rare isotopes, the study of which can provide crucial 
information on short-lived elements normally not found on earth.\cite{FRIB}
Thus, this new experimental program will have widespread impact, ranging from the origin of elements to the              
evolution of the cosmos. 

It is estimated that 
the design and construction of FRIB will take ten years.\cite{FRIB}  
In the meantime, systematic investigations to determine
the properties of asymmetric nuclear matter are proliferating at existing facilities. 

From the theoretical side, some older studies of IANM can be found in Refs.\cite{BC68,Siem}  
Interactions adjusted to fit properties of finite nuclei, such as those based
on the non-relativistic Skyrme Hartree-Fock theory\cite{B+75} or the relativistic mean field
theory,\cite{ST94} have been used to extract 
 phenomenological EoS.                       
Variational calculations of asymmetric matter have also been reported.\cite{12,APR} 
Fuchs {\it et al.}\cite{FLW} defined a Lorentz invariant functional of the baryon field operators              
to project Dirac-Brueckner nuclear matter results onto the meson-nucleon vertices of an effective
density-dependent field theory. This was then applied to asymmetric matter and finite nuclei 
in Hartree calculations.\cite{HKL}  
Extensive work with IANM has also been reported by Lombardo and collaborators.\cite{Catania1,Catania2}  
Dirac-Brueckner-Hartree-Fock 
calculations of IANM properties were performed by the Oslo group,\cite{Oslo}           
the Idaho group,\cite{AS03} and by Fuchs and collaborators.\cite{Fuchs}
Typically, considerable model dependence is observed among the different EoS of IANM, especially in the
high-density region. 

Asymmetric nuclear matter can be characterized by the neutron density, 
$\rho_n$, and the proton density, $\rho_p$, defined as the number of neutrons or protons per unit of volume. 
In infinite matter, they are obtained by summing the neutron or proton states per volume (up to their respective 
Fermi momenta, $k^{n}_{F}$ or $k^{p}_{F}$) and applying the appropriate degeneracy factor. The result is 
\begin{equation}
  \rho_i =\frac{ (k^{i}_{F})^3}{3 \pi ^2} ,   \label{rhonp}   
\end{equation}
with $i=n$ or $p$. 

It is more convenient to refer to the total density
$\rho = \rho_n + \rho_p$ and the asymmetry (or neutron excess) parameter
$\alpha = \frac{ \rho_n - \rho_p}{\rho}$. 
Clearly, $\alpha$=0 corresponds to symmetric matter and 
$\alpha$=1 to neutron matter.                       
In terms of $\alpha$ and the average Fermi momentum, $k_F$, related to the total density in the usual way, 
\begin{equation}
  \rho =\frac{2 k_F^3}{3 \pi ^2} ,   \label{rho}   
\end{equation}
the neutron and proton Fermi momenta can be expressed as 
\begin{equation}
 k^{n}_{F} = k_F{(1 + \alpha)}^{1/3}            \label{kfn}
\end{equation}
and 
\begin{equation}
 k^{p}_{F} = k_F{(1 - \alpha)}^{1/3} ,            \label{kfp} 
\end{equation}
 respectively.

The energy/particle in IANM can, to a very good degree of approximation, be written as 
\begin{equation}
e(\rho, \alpha) \approx e_0({\rho}) + e_{sym}(\rho)\alpha ^2,   \label{e}                    
\end{equation} 
where the first term is the energy/particle in symmetric matter and 
$e_{sym}$ is known as the symmetry energy. In the Bethe-Weizs{\" a}cker formula for the nuclear binding energy, it represents the amount of binding a nucleus has 
to lose when the numbers of protons and neutrons are unequal.                                             
The symmetry energy is also closely related to 
the neutron $\beta$-decay in dense matter, whose threshold depends on the proton fraction. 
A typical value for $e_{sym}$               
at nuclear matter density ($\rho_0$) is 30 MeV, 
with theoretical predictions spreading approximately between 26 and 35 MeV.
The effect of a term of fourth order in the asymmetry parameter (${\cal O}(\alpha ^4)$) on the bulk properties of neutron stars 
is very small, although it may impact the proton fraction at high density. 
More generally, 
non-quadratic terms are usually associated with isovector pairing, which is a surface effect and thus vanishes
in infinite matter.\cite{Steiner}  

Equation~(\ref{e}) displays a convenient separation between the symmetric and aymmetric parts of the EoS, 
which facilitates the identification of observables that may be sensitive, for instance, mainly to the 
symmetry energy. At this time, groups from GSI,\cite{GSI} MSU,\cite{Tsang} Italy,\cite{Greco} France,\cite{IPN} and 
China\cite{China2,China3,China4} 
are investigating the density dependence of the symmetry energy through heavy-ion collisions. 
From recent results, such as those reported at the 2009 ``International Workshop on Nuclear Dynamics           
in Heavy-ion Reactions and the Symmetry Energy"
(Shanghai, China, August 22-25, 2009), these investigations appear to agree reasonably well on the following parametrization 
of the symmetry energy: 
\begin{equation}
e_{sym}(\rho) = 12.5 \, MeV \Big (\frac{\rho}{\rho_0}\Big )^{2/3} +                       
17.5 \, MeV \Big (\frac{\rho}{\rho_0}\Big )^{\gamma_i},                   \label{es} 
\end{equation} 
where the first term is the kinetic contribution and 
 $\gamma_i$ (the exponent appearing in the potential energy part) is found to be between 0.4 and 1.0. 
Naturally, there are uncertainties associated with all transport models. 
Recent constraints from MSU\cite{Tsang} were extracted from simulations of $^{112}$Sn
and $^{124}$Sn collisions with an Improved Quantum Molecular Dynamics transport model and are 
consistent with isospin diffusion data and the ratio of neutron and proton spectra.

Typically, parametrizations like the one given in Eq.~(\ref{es}) are valid 
at or below the saturation density, $\rho_0$. Efforts to constrain the behavior of the symmetry energy
at higher densities 
are presently being pursued through observables such as $\pi ^-/\pi^+$ ratio, 
$K ^+/K^0$ ratio, neutron/proton differential transverse flow, or nucleon elliptic flow.\cite{Ko09} 

A more detailed discussion of the symmetry energy and the status of its theoretical predictions will be presented in    
Sec.~3.2.3.

\section{The Microscopic Approach}

In this section, we present a discussion of the microscopic approach to nuclear matter, in general, and the DBHF
method, in particular.                                

\subsection{The two-body potential}

By {\it ab initio}  we mean that the starting point of the many-body calculation is a realistic NN interaction which is then applied in the 
nuclear medium without any additional free parameters. 
Thus the first question to be confronted concerns the choice of the ``best" NN interaction. 
As already mentioned in the Introduction, 
after the development of QCD and the understanding of its symmetries,  
chiral effective theories\cite{chi} were developed as a way to respect the 
symmetries of QCD while keeping the degrees of freedom (nucleons and pions) typical of low-energy nuclear physics. However, 
chiral perturbation theory (ChPT)
has definite limitations as far as the range of allowed momenta is concerned. 
For the purpose of applications in dense matter, where higher and higher momenta become involved     
with increasing Fermi momentum, NN potentials based on ChPT are unsuitable.       

Relativistic meson theory is an appropriate framework to deal with the high momenta encountered in dense
matter. In particular, 
the one-boson-exchange (OBE) model has proven very successful in describing NN data in free space 
and has a good theoretical foundation. 
Among the many available OBE potentials, some being part of the ``high-precision generation",\cite{pot1,pot2,pot3} 
we seek a momentum-space potential developed within a relativistic scattering equation, such as the 
one obtained through the Thompson\cite{Thom} three-dimensional reduction of the Bethe-Salpeter equation.\cite{BS}
Furthermore, we require a potential that uses 
the pseudovector coupling for the interaction of nucleons with pseudoscalar mesons. 
With these constraints in mind, 
as well as the requirement of a good description of the NN data, 
Bonn B\cite{Mac89} is a reasonable choice. As is well known, the NN potential model dependence
of nuclear matter predictions is not negligible. The saturation points obtained with different NN potentials
move along the famous ``Coester band" depending on the strength of the tensor force, with the weakest tensor
force yielding the largest attraction. This can be understood in terms of medium effects (particularly 
Pauli blocking) reducing the (attractive) second-order term in the expansion of the reaction matrix. 
A large second-order term will undergo a large reduction in the medium. Therefore, noticing that the second-order term
is dominated by the tensor component of  the force, nuclear potentials with a strong tensor component will
yield less attraction in the medium. 
For the same reason (that is, the role of the tensor force in  
nuclear matter), 
the potential model dependence is strongly reduced in pure (or nearly pure) neutron matter, due to the  
absence of isospin-zero partial waves. 

\begin{figure}
\centering            
\vspace*{-3.2cm}
\hspace*{-2.0cm}
\scalebox{1.0}{\includegraphics{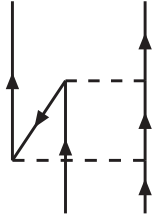}}
\vspace*{-21.0cm}
\caption{Contribution to the NN interaction from virtual pair excitation.                   
Upward- and downward-pointing arrows represent nucleons and antinucleons, respectively.
Dashed lines denote mesons.                            
} 
\label{2b}
\end{figure}

Already when QCD (and its symmetries) were unknown, it was observed that the contribution from the
nucleon-antinucleon pair diagram, Fig.~\ref{2b}, becomes unreasonably large if the pseudoscalar (ps) coupling is used, 
leading to very large pion-nucleon scattering lengths.\cite{GB79}                                             
We recall that the Lagrangian density for pseudoscalar coupling of the nucleon field ($\psi$) with the  pseudoscalar meson
field ($\phi$) is 
\begin{equation}
{\cal L}_{ps} = -ig_{ps}\bar {\psi} \gamma _5 \psi \phi.     \label{ps} 
\end{equation} 
On the other hand, the same contribution (Fig.~\ref{2b})
is heavily reduced by the pseudovector (pv) coupling (a mechanism which
became known as ``pair suppression"). The reason for the suppression is the presence of the 
covariant derivative                                                                                     
at the pseudovector vertex,                                                  
\begin{equation}
{\cal L}_{pv} = \frac{f_{ps}}{m_{ps}}{\bar \psi}  \gamma _5 \gamma^{\mu}\psi \partial_{\mu} \phi,              
\label{pv} 
\end{equation} 
which reduces the contribution of the vertex for low momenta and, thus, 
 explains the small value of the pion-nucleon
scattering length at threshold.\cite{GB79}  
Considerations based on chiral symmetry\cite{GB79} can further motivate 
the choice of the pseudovector coupling.                          

In closing this section, we wish to highlight          
the most important aspect of the ``{\it ab initio}" approach: namely, the only free parameters of the
model (the parameters of the NN potential)                                               
are determined by fitting the free-space NN data and never readjusted in the medium. In other
words, the model parameters are tightly constrained and the calculation in the medium is 
parameter free. 
The presence of free parameters in the medium would generate effects and sensitivities which are hard to
control and interfere with the predictive power of the theory. 

\subsection{The Dirac-Brueckner-Hartree-Fock approach to symmetric and  asymmetric nuclear matter}

\subsubsection{Formalism} 

\begin{figure}[!t] 
\centering         
\vspace*{-3.2cm}
\hspace*{-2.0cm}
\scalebox{0.9}{\includegraphics{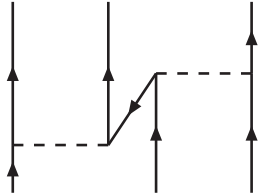}}
\vspace*{-19.0cm}
\caption{Three-body force due to virtual pair excitation. Conventions as in the previous figure.
} 
\label{3b}
\end{figure}

The main strength of the DBHF approach is in its inherent ability to account for important three-body forces   
through its density dependence. 
In Fig.~\ref{3b} we show a three-body force (TBF) originating from virtual excitation of a nucleon-antinucleon pair, 
known as ``Z-diagram". Notice that the observations from the previous section ensure that the corresponding diagram
at the two-body level, Fig.~\ref{2b}, is moderate in size when the pv coupling is used. 
The main feature of                                     
the DBHF method turns out to be closely related to 
the TBF depicted in Fig.~\ref{3b}, as we will argue next. In the DBHF approach, one describes the positive energy solutions
of the Dirac equation in the medium as 
\begin{equation}
u^*(p,\lambda) = \left (\frac{E^*_p+m^*}{2m^*}\right )^{1/2}
\left( \begin{array}{c}
 {\bf 1} \\
\frac{\sigma \cdot \vec {p}}{E^*_p+m^*} 
\end{array} 
\right) \;
\chi_{\lambda},
\label{ustar}
\end{equation}
where the effective mass, $m^*$, is defined as $m^* = m+U_S$, with $U_S$ an attractive scalar potential.
(This will be derived below.) 
It can be shown that both the description of a single-nucleon via Eq.~(\ref{ustar}) and the evaluation of the 
Z-diagram, Fig.~\ref{3b}, generate a repulsive effect on the energy/particle in symmetric nuclear matter which depends on the density approximately
as 
\begin{equation}
\Delta E \propto  \left (\frac{\rho}{\rho_0}\right )^{8/3} \, , 
\label{delE} 
\end {equation}
and provides the saturating mechanism missing from conventional Brueckner calculations. 
(Alternatively, explicit TBF are used along with the BHF method in order to achieve a similar result.) 
Brown showed that the bulk of the desired effect can be obtained as a lowest order (in $p^2/m$) relativistic correction
to the single-particle propagation.\cite{GB87}  
With the in-medium spinor as in Eq.~(\ref{ustar}), the correction to the free-space spinor can be written 
approximately as 
\begin{equation}
u^*(p,\lambda) -u(p,\lambda)\approx                                                  
\left( \begin{array}{c}
 {\bf 0} \\
-\frac{\sigma \cdot \vec {p}}{2 m^2}U_S
\end{array} 
\right) \;
\chi_{\lambda},
\label{delu} 
\end{equation}
where for simplicity the spinor normalization factor has been set equal to 1, in which case it 
is clearly seen that the entire effect originates from the modification of the spinor's lower component. 
By expanding the single-particle energy to order $U_S^2$, Brown showed that the correction to the 
energy consistent with Eq.~(\ref{delu}) can be written as $\frac{p^2}{2m}(\frac{U_S}{m})^2$. He then proceeded to 
estimate the correction to the energy/particle and found it to be approximately as given in Eq.~(\ref{delE}). 

The approximate equivalence of the effective-mass description of Dirac states and the contribution from the Z-diagram 
has a simple intuitive explanation in the observation 
that Eq.~(\ref{ustar}), like any other solution of the Dirac equation,
can be written as a superposition of positive and negative energy solutions. On the other hand, the ``nucleon" in the 
middle of the Z-diagram, Fig.~\ref{3b}, is precisely a superposition of positive and negative energy states. 
In summary, the DBHF method effectively takes into account a particular class of 
TBF, which are crucial for nuclear matter saturation. 

\begin{figure}
\begin{center}
\vspace*{-3.0cm}
\hspace*{-2.0cm}
\scalebox{0.9}{\includegraphics{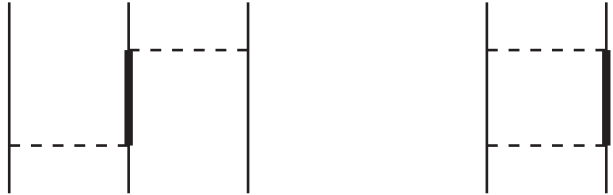}}
\vspace*{-19.0cm}
\caption{Left: three-body force arising from $\Delta$-isobar excitation (thick line). 
Right: two-meson exchange contribution to the NN interaction involving 
$\Delta$-isobar excitation.
} 
\label{Delta}
\end{center}
\end{figure}

Of course, 
other, more popular, three-body forces (not included in DBHF) need to be addressed as well.   
Figure~\ref{Delta} shows the TBF that is included in essentially all TBF models, regardless
other components; it is the Fujita-Miyazawa TBF.\cite{FM}                                           
With the addition of contributions from $\pi N$ S-waves, one ends up with the 
well-known Tucson-Melbourne TBF.\cite{TM} The microscopic TBF of Ref.\cite{Catania3} includes 
contributions from excitation of the Roper resonance (P$_{11}$ isobar) as well. 

Now, if diagrams such as the one shown on the left-hand side of  Fig.~\ref{Delta} are taken into account, 
consistency requires that medium modifications at the corresponding two-body level are also included, 
that is, the diagram on the right-hand side of Fig.~\ref{Delta} should be present and properly medium modified. 
Large cancellations then take place, a fact that was brought up a long time ago.\cite{DMF}                        
When the two-body sector is handled via OBE diagrams, the two-pion exchange is 
effectively incorporated through the $\sigma$ ``meson", which 
cannot generate the (large) medium effects (dispersion and Pauli blocking on $\Delta$
intermediate states) required by the consistency arguments presented above. Thus, caution needs to
be exercised when applying TBF in a particular space (of nucleons only, or nucleons and $\Delta$).

Having first summarized the main DBHF philosophy, 
we now proceed to describe the DBHF calculation of IANM.\cite{AS03}
In the end, this will take us back to the crucial point of the DBHF approximation, Eq.~(\ref{ustar}). 

We start from the Thompson\cite{Thom} relativistic three-dimensional reduction 
of the Bethe-Salpeter equation.\cite{BS} The Thompson equation is applied to nuclear matter in
strict analogy to free-space scattering and reads, in the nuclear matter rest frame,                 
\begin{eqnarray}
&& g_{ij}(\vec q',\vec q,\vec P,(\epsilon ^*_{ij})_0) = v_{ij}^*(\vec q',\vec q) \nonumber \\            
&& + \int \frac{d^3K}{(2\pi)^3}v^*_{ij}(\vec q',\vec K)\frac{m^*_i m^*_j}{E^*_i E^*_j}
\frac{Q_{ij}(\vec K,\vec P)}{(\epsilon ^*_{ij})_0 -\epsilon ^*_{ij}(\vec P,\vec K)} 
g_{ij}(\vec K,\vec q,\vec P,(\epsilon^*_{ij})_0) \, ,                                   
\label{gij}
\end{eqnarray}                    
where $g_{ij}$ is the in-medium reaction matrix 
($ij$=$nn$, $pp$, or $np$), and the                                      
asterix signifies that medium effects are applied to those quantities. Thus the NN potential, 
$v_{ij}^*$, is constructed in terms of effective Dirac states (in-medium spinors) as explained above. 
In Eq.~(\ref{gij}),                                  
$\vec q$, $\vec q'$, and $\vec K$ are the initial, final, and intermediate
relative momenta, and $E^*_i = \sqrt{(m^*_i)^2 + K^2}$. 
The momenta of the two interacting particles in the nuclear matter rest frame have been expressed in terms of their
relative momentum and the center-of-mass momentum, $\vec P$, through
\begin{equation} 
\vec P = \vec k_{1} + \vec k_{2} \, ,  \label{P}    
\end{equation} 
\begin{equation} 
\vec K = \frac{\vec k_{1} - \vec k_{2}}{2} \, .  \label{K}
\end{equation}                    
The energy of the two-particle system is 
\begin{equation} 
\epsilon ^*_{ij}(\vec P, \vec K) = 
e^*_{i}(\vec P, \vec K)+  
e^*_{j}(\vec P, \vec K)   
\label{eij}
\end{equation} 
 and $(\epsilon ^*_{ij})_0$ is the starting energy.
 The single-particle energy $e_i^*$ includes kinetic energy and potential 
 energy contributions (see Eq.~(\ref{spe}) below).                               
The Pauli operator, $Q_{ij}$, prevents scattering to occupied $nn$, $pp$, or $np$ states.            
 To eliminate the angular
dependence from the kernel of Eq.~(\ref{gij}), it is customary to replace the exact
Pauli operator with its angle-average. 
Detailed expressions for the Pauli operator                     
and the average center-of-mass momentum in the case of two different Fermi seas  
can be found in Ref.\cite{AS03}.                              

With the definitions
\begin{equation} 
G_{ij} = \frac{m^*_i}{E_i^*(\vec{q'})}g_{ij}
 \frac{m^*_j}{E_j^*(\vec{q})}             
\label{Gij}
\end{equation} 
and 
\begin{equation} 
V_{ij}^* = \frac{m^*_i}{E_i^*(\vec{q'})}v_{ij}^*
 \frac{m^*_j}{E_j^*(\vec{q})} \, ,        
\label{Vij}
\end{equation} 
 one can rewrite Eq.~(\ref{gij}) as
\begin{eqnarray}
&& G_{ij}(\vec q',\vec q,\vec P,(\epsilon ^*_{ij})_0) = V_{ij}^*(\vec q',\vec q) \nonumber \\[4pt]
&& + \int \frac{d^3K}{(2\pi)^3}V^*_{ij}(\vec q',\vec K)
\frac{Q_{ij}(\vec K,\vec P)}{(\epsilon ^*_{ij})_0 -\epsilon ^*_{ij}(\vec P,\vec K)} 
G_{ij}(\vec K,\vec q,\vec P,(\epsilon^*_{ij})_0) \, ,                                    
\label{Geq}
\end{eqnarray}                    
which is formally identical to its non-relativistic counterpart.

The goal is to determine self-consistently the nuclear matter single-particle potential   
which, for IANM, will be different for neutrons and protons. 
To facilitate the description of the procedure, we will use a schematic
notation for the neutron/proton potential.                                                   
We write, for neutrons,
\begin{equation}
U_n = U_{np} + U_{nn} \; , 
\label{un}
\end{equation}
and for protons
\begin{equation}
U_p = U_{pn} + U_{pp} \, , 
\label{up}
\end{equation}
where each of the four pieces on the right-hand-side of Eqs.~(\ref{un}-\ref{up}) signifies an integral of the appropriate 
$G$-matrix ($nn$, $pp$, or $np$) obtained from Eq.~(\ref{Geq}).                                           
Clearly, the two equations above are coupled through 
the $np$ component and so they must be solved simultaneously. Furthermore, 
the $G$-matrix equation and Eqs.~(\ref{un}-\ref{up})  
are coupled through the single-particle energy (which includes the single-particle
potential, itself defined in terms of the $G$-matrix). So we have a coupled system to be solved self-consistently.

Before proceeding with the self-consistency, 
one needs an {\it ansatz} for the single-particle potential. The latter is suggested by 
the most general structure of the nucleon self-energy operator consistent with 
all symmetry requirements. That is: 
\begin{equation}
{\cal U}_i({\vec p}) =  U_{S,i}(p) + \gamma_0  U_{V,i}^{0}(p) - {\bf \gamma}\cdot {\vec p}  U_{V,i}(p) \, , 
\label{Ui1}
\end{equation}
where $U_{S,i}$ and 
$U_{V,i}$ are an attractive scalar field and a repulsive vector field, respectively, with 
$ U_{V,i}^{0}$ the timelike component of the vector field. These fields are in general density and momentum dependent. 
We take             
\begin{equation}
{\cal U}_i({\vec p}) \approx U_{S,i}(p) + \gamma_0 U_{V,i}^{0}(p) \, ,                                            
\label{Ui2}
\end{equation}
which amounts to assuming that the spacelike component of the vector field is much smaller than 
 both $U_{S,i}$ and $U_{V,i}^0$. Furthermore, neglecting the momentum dependence of the scalar and
vector fields and inserting Eq.~(\ref{Ui2}) in the Dirac equation for neutrons/protons propagating in 
nuclear matter,
\begin{equation}
(\gamma _{\mu}p^{\mu} - m_i - {\cal U}_i({\vec p})) u_i({\vec p},\lambda) = 0  \, ,                                                       
\label{Dirac1} 
\end{equation}
naturally leads to rewriting the Dirac equation in the form 
\begin{equation}
(\gamma _{\mu}p^{\mu *} - m_i^*) u_i({\vec p},\lambda) = 0  \, ,                                                       
\label{Dirac2} 
\end{equation}
with positive energy solutions as in Eq.~(\ref{ustar}), $m_i^* = m + U_{S,i}$, and 
\begin{equation}
(p^0)^* = p^0 - U_{V,i}^0 (p) \, .                                                                 
\label{p0}
\end{equation}
The subscript ``$i$'' signifies that these parameters are different for protons and
neutrons. 

As in the symmetric matter case,\cite{BM84} evaluating  the expectation value of Eq.~(\ref{Ui2})       
leads to a parametrization of 
the single particle potential for protons and neutrons (Eqs.(\ref{un}-\ref{up})) in terms of the 
constants $U_{S,i}$ and $U_{V,i}^0$ which is given by      
\begin{equation}
U_i(p) = \frac{m^*_i}{E^*_i}<{\vec p}|{\cal U}_i({\vec p})|{\bf p}> = 
\frac{m^*_i}{E^*_i}U_{S,i} + U_{V,i}^0 \; .      
\label{Ui3}
\end{equation}
Also, 
\begin{equation}
U_i(p) =                                                              
\sum_{p'_j \le k_F^i} G_{ij}({\vec p} _i,{\vec p}'_j) \; , 
\label{Ui4}
\end{equation}
which, along with Eq.~(\ref{Ui3}), allows the self-consistent determination of the single-particle
potential as explained below. 

The kinetic contribution to the single-particle energy is
\begin{equation}
T_i(p) = \frac{m^*_i}{E^*_i}<{\vec p}|\gamma \cdot {\vec p} + m|{\vec p}> =     
\frac{m_i m^*_i + {\vec p}^2}{E^*_i} \; , 
\label{KE}    
\end{equation}
and the single-particle energy is 
\begin{equation}
e^*_i(p) = T_i(p) + U_i(p) = E^*_i + U^0_{V,i} \; . 
\label{spe}
\end{equation}
The constants $m_i^*$ and 
\begin{equation}
U_{0,i} = U_{S,i} + U_{V,i}^0      
\label{U0i} 
\end{equation}
are convenient to work with as they 
facilitate          
the connection with the usual non-relativistic framework.\cite{HT70}                        

Starting from some initial values of $m^*_i$ and $U_{0,i}$, the $G$-matrix equation is 
 solved and a first approximation for $U_{i}(p)$ is obtained by integrating the $G$-matrix 
over the appropriate Fermi sea, see Eq.~(\ref{Ui4}). This solution is 
again parametrized in terms of a new set of constants, determined by fitting the parametrized $U_i$, 
Eq.~(\ref{Ui3}), 
to its values calculated at two momenta, a procedure known as the ``reference spectrum approximation". 
The iterative procedure is repeated until convergence is reached.     

Finally, the energy per neutron or proton in nuclear matter is calculated from 
\begin{equation}
\bar{e}_{i} = \frac{1}{A}<T_{i}> + \frac{1}{2A}<U_{i}> -m \; . 
\label{ei}
\end{equation}
 The EoS, or energy per nucleon as a function of density, is then written as
\begin{equation}
    \bar{e}(\rho_n,\rho_p) = \frac{\rho_n \bar{e}_n + \rho_p \bar{e}_p}{\rho} \, , 
\label{enp} 
\end{equation}
or 
\begin{equation}
    \bar{e}(k_F,\alpha) = \frac{(1 + \alpha) \bar{e}_n + (1-\alpha) \bar{e}_p}{2} \, . 
\label{eav} 
\end{equation}
Clearly, symmetric nuclear matter is obtained as a by-product of the calculation described above 
by setting $\alpha$=0. 

In the DBHF calculation of Ref.\cite{Fuchs}, 
a similar scheme is applied to obtain the self-consistent $G$-matrix and spinor basis. At each 
step of the iterative procedure, the nucleon self-energy is first calculated by integrating the $G$-matrix 
elements over the Fermi sea. 
The self-energy components are then obtained from 
the appropriate traces\cite{Fuchs}, i.~e. 
\begin{equation}
\Sigma _S = \frac{1}{4} tr[\Sigma] \, , \, \, \, 
\Sigma _0 = -\frac{1}{4} tr[\gamma _0\Sigma] \, , \, \, \, 
\Sigma _V = -\frac{1}{4|{\vec p}|^2} tr[\gamma \cdot {\vec p}\Sigma] \, . 
\label{Sig}
\end{equation}
A covariant representation of the $G$-matrix is used to facilitate the transitions between the two-nucleon 
center-of-mass frame and the nuclear matter rest frame. 
Again, the presence of the medium naturally leads to define an effective mass, which                  
is given by\cite{Fuchs} 
\begin{equation}
m^*(p,k_F) = (m + Re \Sigma _S(p,k_F))/(1 + \Sigma _V(p,k_F)) \, . 
\label{mstar} 
\end{equation}
Although the effective mass is density and momentum dependent, ultimately it is defined as the
value of the expression given in Eq.~(\ref{mstar}) at $p=k_F$. Thus, the ``reference spectrum 
approximation" is also employed in Ref.\cite{Fuchs}, although differently                          
in some of the technical aspects as compared to the Idaho method. 
These technical aspects appear to impact the predictions of some isovector quantities, as it will be
discussed in Sec.~3.2.4.

\subsubsection{EoS predictions with the DBHF approach} 

\begin{figure}
\begin{center}
\vspace*{-1.0cm}
\hspace*{-2.0cm}
\scalebox{0.3}{\includegraphics{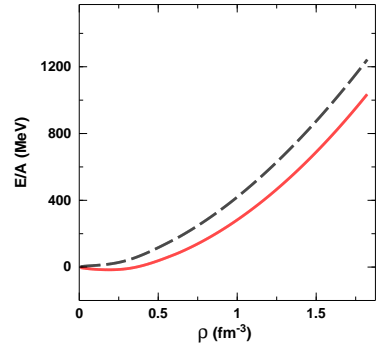}}
\vspace*{-2.5cm}
\caption{DBHF predictions for the EoS of symmetric matter (solid red) and neutron matter (dashed black).            
} 
\label{eos}
\end{center}
\end{figure}

In Fig.~\ref{eos}, we show EoS predictions for symmetric matter (solid red) and neutron matter (dashed black)  
as obtained from the Idaho calculation described in the previous section.
Equation~(4) then gives the EoS values for any $\alpha$, a behavior which can be verified to be       
approximately true.\cite{AS03}                                                  

The EoS from DBHF can be characterized as being moderately soft at low to medium density                      
and  fairly ``stiff" at 
high densities.                                                                                         
The predicted saturation density and energy for the symmetric matter EoS in Fig.~\ref{eos} are equal to 0.185 fm$^{-3}$ and -16.14 MeV, respectively, 
and the compression modulus is 252 MeV. For comparison, the same saturation obserbables as 
predicted by the other DBHF model presently on the market\cite{Fuchs} are 0.181 fm$^{-3}$ and -16.15 MeV
for saturation density and energy, and  230 MeV for the incompressibility. 

The increased stiffness featured by the DBHF EoS at the higher densities 
originates from the strongly density-dependent repulsion inherent to the 
Dirac-Brueckner-Hartee-Fock method.  
In Ref.\cite{Fuchs2}, it is pointed out 
that constraints from neutron star phenomenology together with flow data from heavy-ion           
reactions suggest that such EoS behavior may be desirable. 
We will come back to this point later, in conjunction with neutron star predictions. 

\begin{figure}
\begin{center}
\scalebox{0.28}{\includegraphics[angle=-90]{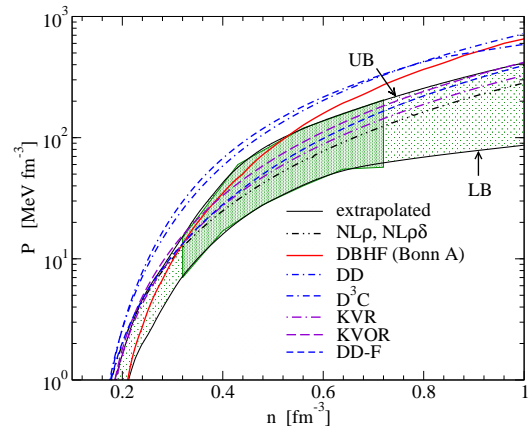}}
\vspace*{0.6cm}
\caption{Pressure in symmetric matter predicetd by various models. The shaded area corresponds to the region
of pressure consistent with the flow data analysed in Ref.$^{56}$. [Figure reprinted with permission
from T. Kl{\"a}hn.$^{51}$ Copyright (2006) by the American Physical Society.      
http://prc.aps.org/abstract/PRC/v74/i3/e035802]
}
\label{Klaen}
\end{center}
\end{figure}

At this point, it it useful 
to take a closer look at 
some of the predictions included in the analysis of Ref.\cite{Fuchs2}, such as 
relativistic mean field (RMF) models. Examples are those from 
 Refs.\cite{DD,D3C}, which use density-dependent (``DD") meson couplings and are 
fitted to the properties of nuclei up to about 0.15 fm$^{-3}$. They generate the steepest 
EoS and thus the largest pressures, see Fig.~\ref{Klaen}.                                                
An improvement to the traditional RMF description of nuclear matter 
can be obtained through the introduction of non-linear (``NL") self-interactions of the $\sigma$ meson, such as done 
in the models of Refs.\cite{NL1,NL2}, with the parametrization of Ref.\cite{NL2} including 
both the $\delta$ and the $\rho$ mesons in the isovector channel. The corresponding EoS are much less repulsive
than those of ``DD" models (although the symmetry energy becomes very large at high 
density, possibly due to the absence of non-linearity and density dependence at the 
isovector level). 

\begin{figure}
\begin{center}
\vspace*{-2.0cm}
\hspace*{-2.0cm}
\scalebox{0.4}{\includegraphics{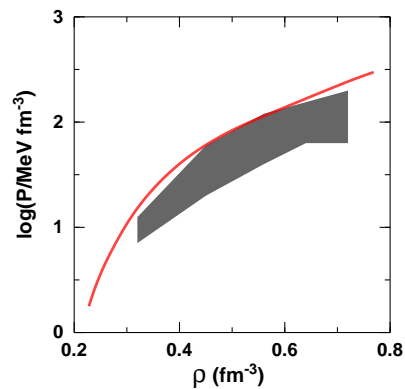}}
\vspace*{-2.5cm}
\caption{Pressure in symmetric matter from the Idaho DBHF calculation. The shaded area corresponds to the region
of pressure consistent with the flow data analysed in Ref.$^{56}$.}                  
\label{psm}
\end{center}
\end{figure}

Clearly, the pressure as a function of density plays the crucial role in building the structure of    
a neutron star. 
In Fig.~\ref{psm} we show the pressure in symmetric matter as predicted by the Idaho calculation compared with constraints obtained
from flow data.\cite{MSU}  
The predictions are seen to fall just on the high side of the constraints and grow
rather steep at high density.                     
Comparing with Fig.~\ref{Klaen}, 
 we see that                                                      
the Idaho predictions                                                                      
are well below those of DD-RMF models at low to 
moderate density but nearly catch up with them at very high density, a description that    
would also be appropriate for the 
DBHF predictions of Ref.\cite{Fuchs} (red curve in Fig.~\ref{Klaen}).                                            
Of all the cases studied in 
Ref.\cite{Fuchs2}, DD-RMF models predict the largest maximum masses and radii and the 
lowest central densities. Thus,           
an equation of state where high pressure is sustained for a longer radial distance (moving away 
from the center of the star) 
will allow the maximum mass star to be heavier, larger, and more ``diffuse" at the center. 
On the other hand, 
microscopic relativistic models, (such as the DBHF calculation of Ref.\cite{Fuchs}
or the present one, which are in fair agreement with each other), display a rather different              
density dependence of the pressure and produce smaller and more compact maximum mass stars. 
(All other EoS considered in Ref.\cite{Fuchs2} are softer and generate smaller maximum masses 
with smaller radii and larger central densities.)                           

\begin{figure}
\begin{center}
\vspace*{-2.0cm}
\hspace*{-2.0cm}
\scalebox{0.4}{\includegraphics{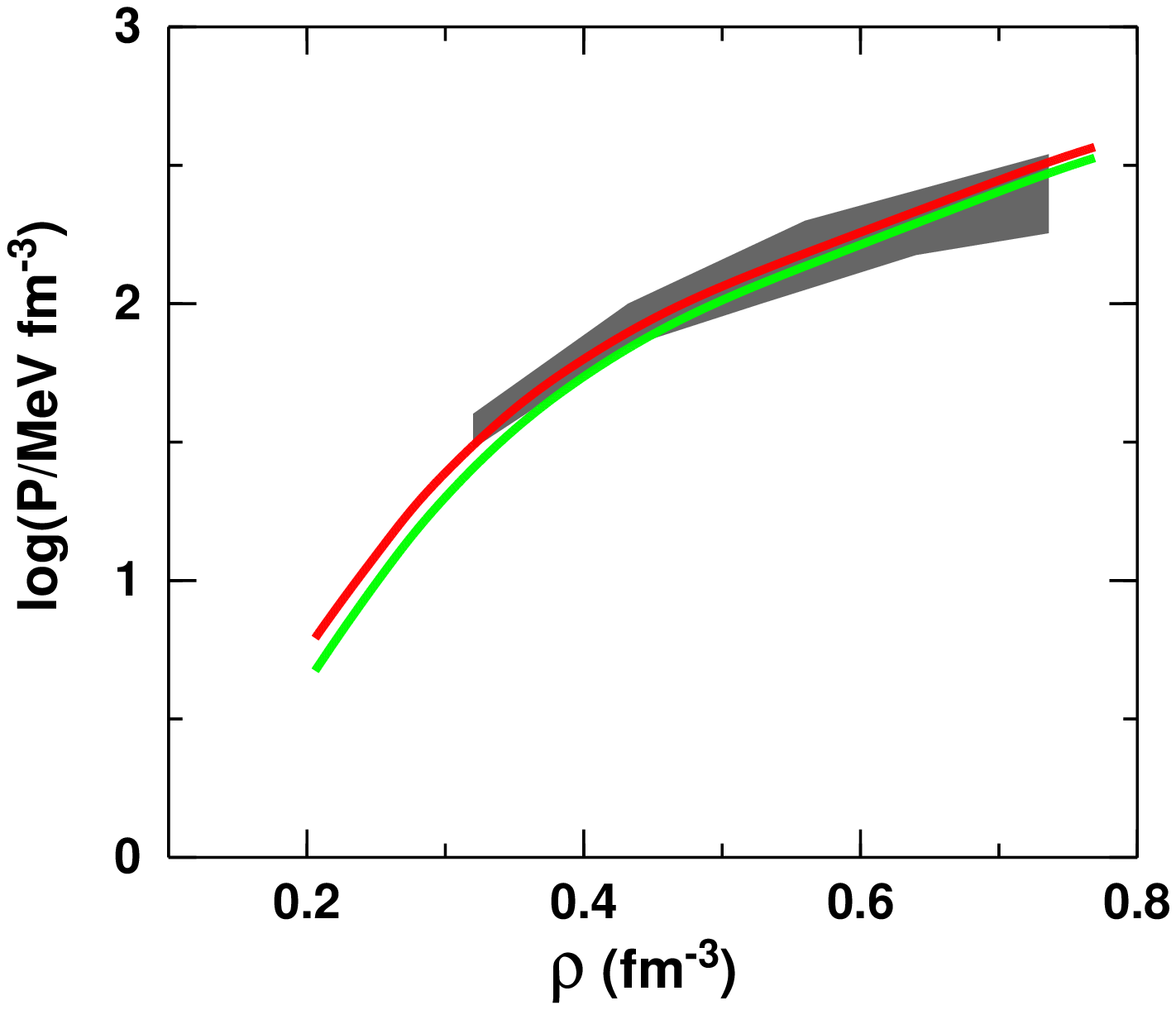}}
\vspace*{-2.5cm}
\caption{Pressure in neutron (red curve) and baryon-lepton (green curve) matter from the Idaho DBHF calculation. The
shaded area corresponds to the region
of pressure consistent with flow data and the inclusion of strong                            
density dependence in the asymmetry terms.$^{56}$}                    
\label{pnm}
\end{center}
\end{figure}

To conclude this section, we show in Fig.~\ref{pnm} the pressure in neutron matter (red curve)
and $\beta$-equilibrated matter (green) as predicted by DBHF calculations. The pressure contour is again from Ref.\cite{MSU} 
and was obtained from flow data together with the assumption of strong density dependence in 
the asymmetry term (indicated as ``Asy\_ stiff'' in Ref.\cite{MSU}).

\subsubsection{The symmetry energy and related observables}

Here we will focus specifically on the symmetry energy and its impact on 
the structure of neutron-rich nuclei, in particular the neutron skin thickness.

\begin{figure}
\begin{center}
\vspace*{-1.0cm}
\hspace*{-2.0cm}
\scalebox{0.4}{\includegraphics{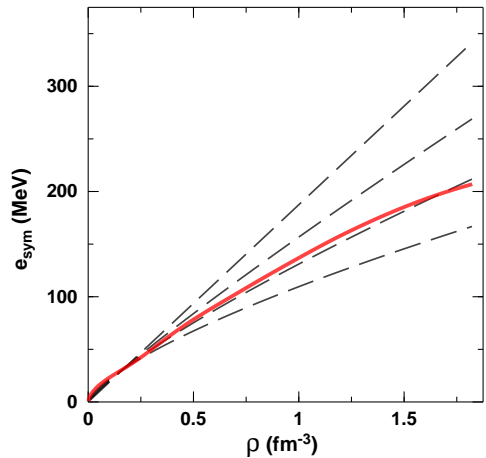}}
\vspace*{-3.0cm}
\caption{DBHF prediction for the symmetry energy (solid red) compared with various         
phenomenological parametrizations (dashed black). See text for details. 
} 
\label{esym}
\end{center}
\end{figure}

In Fig.~\ref{esym}, we display the Idaho DBHF prediction for the symmetry energy by the solid red curve. 
The curve is seen to grow at a lesser rate with increasing density, 
 an indication that, at large density,   
repulsion in the symmetric matter EoS increases more rapidly than 
in the neutron matter EoS.   
This can be understood in terms of increased repulsion in isospin zero partial waves (absent
from neutron matter) as a function of density.                       
Our predicted value for the symmetry pressure, $L$, (see Eq.~(\ref{L}) below), is 69.6 MeV. 

The various black dashed curves in Fig.~\ref{eos} are obtained with the simple parametrization               
\begin{equation}
e_{sym}=C(\rho/\rho_0)^{\gamma} \, ,  
\label{esymm} 
\end{equation}
with $\gamma$ increasing from 0.7 to 1.0 in steps of 0.1, and $C \approx 32$ MeV. 
It seems that a value of $\gamma$ close to 0.8 gives a reasonable description of the DBHF predictions, 
although the use of different functions in different density regions would be best for an 
optimal fit.                                                                                            

Considering that all of the dashed curves are 
commonly used parametrizations 
suggested by heavy-ion data,\cite{BA}  
Fig.~\ref{eos} clearly reflects our limited knowledge of the symmetry energy,      
particularly, but not exclusively, at the larger densities.

As already mentioned in Sec.~2,               
from the experimental side intense effort is going on to obtain reliable empirical information for the less
known aspects of the EoS. Heavy-ion reactions are a popular way to seek constraints on the symmetry 
energy, through analyses of observables that are sensitive to the pressure gradient between 
nuclear and neutron matter. 
Isospin diffusion data from heavy-ion collisions, together with analyses based on isospin-dependent transport
models, can provide information on the slope of the symmetry energy.                         

Concerning the lower densities,     
isospin-sensitive observables can also be identified among the properties of normal nuclei. 
The neutron skin of neutron-rich nuclei is a powerful isovector observable, being sensitive to the   
slope of the symmetry energy, which determines to which extent neutrons will tend to spread outwards 
 to form the skin. 
It is the purpose of this section to systematically examine and discuss the symmetry energy properties 
 in microscopic models and the corresponding neutron skin predictions.  We will take the skin of
$^{208}$Pb as our representative isovector ``observable".
                           
Parity-violating electron scattering experiments are now a realistic option        
to determine neutron distributions with unprecedented accuracy. The neutron radius of 
$^{208}$Pb is expected to be measured within 0.05 fm thanks to the electroweak program
at the Jefferson Laboratory.\cite{Piek06} This level of accuracy could not be achieved with hadronic scattering. 
Parity-violating electron scattering at low momentum transfer is especially suitable to probe neutron densities, as the     
 $Z^0$ boson couples primarily to neutrons. 
With the success of this program, 
 reliable empirical information on neutron skins will be able to provide, in turn, much needed {\it independent} constraint on the 
density dependence of the symmetry energy.

We form an energy functional based on the semi-empirical mass formula, where the volume and  
symmetry terms are contained in the isospin-asymmetric equation of state. Thus, we write the 
energy of a (spherical) nucleus as 
\begin{eqnarray}      
E(Z,A) = \int d^3 r~ e(\rho(r),\alpha(r))\rho(r) + \nonumber  \\ 
+ \int d^3 r f_0(|\nabla \rho(r)|^2 +                       
 \beta|\nabla \rho_I(r)|^2) + Coulomb \, \,  term \, . 
\label{edrop}
\end{eqnarray}      
In the above equation, 
$\rho$ and $\rho_I$ are the usual isoscalar and isovector densities, given by $\rho_n +\rho_p$ and 
$\rho_n -\rho_p$, respectively, in terms of neutron and proton densities. $\alpha$ is the neutron asymmetry
parameter, $\alpha=\rho_I/\rho$, and $e(\rho,\alpha)$ is the energy per particle in 
isospin-asymmetric nuclear matter. 
The latest Idaho EoS\cite{Sam0806} will be used in Eq.~(\ref{edrop}), along with others from different models.

From fits to nuclear binding energies, 
the constant $f_0$ in Eq.~(\ref{edrop}) is approximately 70 $MeV$ $fm^5$, whereas the contribution of the term proportional to $\beta$
was found to be minor.\cite{Oya98} Thus it is reasonable to neglect it.           
(The magnitude of  $\beta$ was estimated to be about 1/4 in Ref.\cite{Furn}, where it was observed that, 
even with variations of $\beta$ between -1 and +1, the effect of the $\beta$ term on the neutron skin 
was negligibly small.) 

The parameters of the proton and neutron densities are obtained by minimizing the value
of the energy functional, Eq.~(\ref{edrop}), assuming Thomas-Fermi distribution functions. 
Although simple, this method has the advantage of allowing a         
very direct connection between the EoS and the properties of finite nuclei. (It could be used, for 
instance, to determine a semi-phenomenological EoS by fitting it to both binding energies and
charge radii of closed-shell nuclei.) 

\begin{figure}[!t]
\centering           
\includegraphics[totalheight=2.0in]{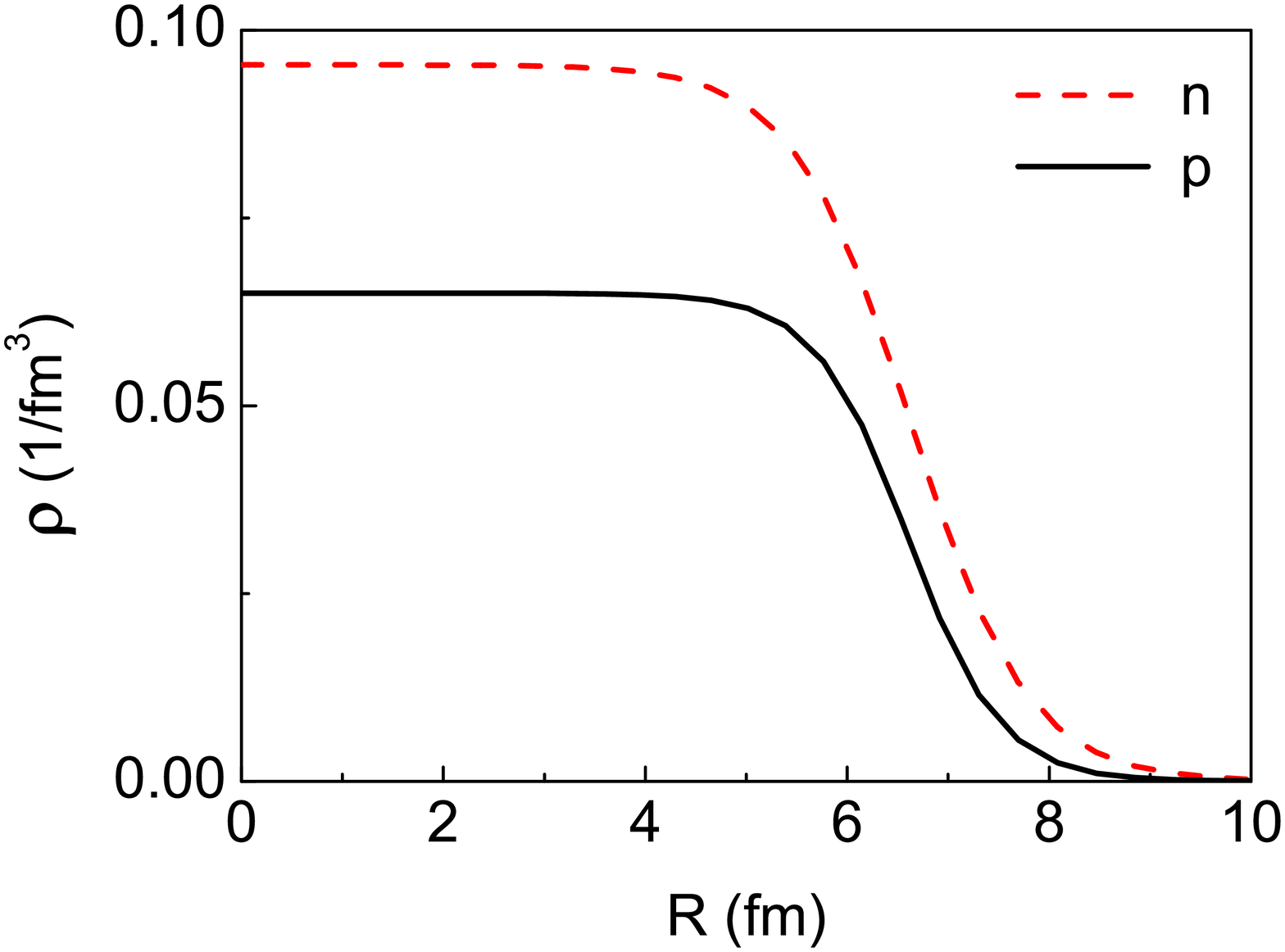} 
\vspace*{0.2cm}
\caption{(color online) Neutron (red) and proton (black) point densities as obtained from the 
DBHF model.                          
} 
\label{lead}
\end{figure}

In Fig.~\ref{lead}, we show the proton and neutron Thomas-Fermi distributions for $^{208}$Pb as obtained with the method described above and the DBHF model for the       
EoS. The predicted proton and neutron root-mean-square radii are 
 5.39 fm and 5.56 fm, respectively.                                                                      

Typically, 
 predictions of the symmetry energy at saturation density encountered in the literature are in reasonable agreement with 
one another, ranging approximately from 26 to 35 MeV. On the other hand, 
the slope of the symmetry energy is more model dependent and less constrained.     
This is seen through the symmetry pressure, defined as 
\begin{equation}
L = 3 \rho_0 \Big (\frac{\partial e_{sym}(\rho)}{\partial \rho}\Big )_{\rho_0} \approx 
 3 \rho_0 \Big (\frac{\partial e_{n.m.}(\rho)}{\partial \rho}\Big )_{\rho_0} \, . 
\label{L} 
\end{equation} 
Thus, $L$ is sensitive to the gradient of the energy per particle in neutron matter ($e_{n.m.}$). 
As to be expected on physical grounds, the neutron skin, given by                                  
\begin{equation}
S = \sqrt{<r_n^2>} - \sqrt{<r_p^2>} \, \, , 
\label{S} 
\end{equation} 
is highly sensitive to the same gradient.

Values of $L$ are reported to range 
from -50 to 100 MeV as seen, for instance, through the numerous
parametrizations of Skyrme interactions (see Ref.\cite{BA} and references therein),               
 all chosen to fit the binding energies and the 
charge radii of a large number of nuclei.  
Heavy-ion data impose boundaries for $L$ at $85 \pm 25$ MeV,\cite{Chen07,Dan07} with more              
stringent constraints being presently extracted.\cite{Tsang} 
Other reports\cite{Ko09} state the constraints as 
$85 \pm 55$ MeV.                                           
Also, a nearly linear 
correlation is observed between the neutron skin $S$ and the $L$               
parameter.\cite{Li05}        

Such phenomenological studies are extremely useful, but, 
ultimately, they must be compared with {\it ab initio} approaches in order to get true        
physical insight.                                                                      
The direct connection with the underlying nuclear forces will then facilitate the physical understanding,
when combined with reliable constraints.

As explained in Sec.~3.2.1,            
the DBHF model does not include three-body forces explicitely, but effectively incorporates the 
class of TBF originating from the presence of nucleons and antinucleons (the          
``Z-diagrams" in Fig.~\ref{3b}),               
see previous discussion.  
As the other main input of our comparison, we will take the EoS from the microscopic approach 
of Ref.\cite{Catania3}. There (and in previous work by the same authors), 
the Brueckner-Hartree-Fock (BHF) formalism is employed along with microscopic three-body forces.        
In particular, in Ref.\cite{Catania4}
the meson-exchange TBF are constructed applying the same parameters
as used in the corresponding nucleon-nucleon (NN) potentials, which are: Argonne V18\cite{V18} (V18), Bonn B\cite{Mac89} (BOB), Nijmegen 93\cite{N93} (N93). 
The popular (but phenomenological) Urbana TBF\cite{UIX} (UIX) is also utilized in Ref.\cite{Catania4}. Convenient
parametrizations in terms of simple analytic functions are given in all cases and we
will use those to generate the various EoS. We will refer to this approach, generally, as ``BHF + TBF". 
This comparison will bring up the discussion on microscopic models and explicit TBF
outlined in Sec.~3.2.

\begin{figure}[!t] 
\centering          
\includegraphics[totalheight=2.0in]{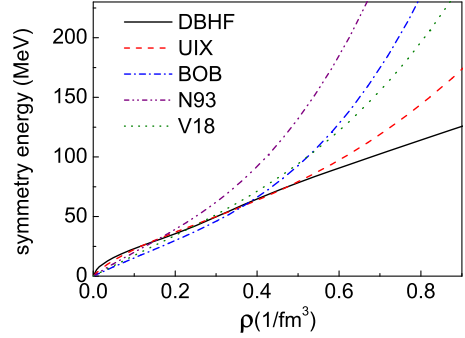} 
\vspace*{-0.3cm}
\caption{(color online) Predictions for the symmetry energy from DBHF and various
``BHF + TBF" models discussed in the text. 
} 
\label{esmicro}
\end{figure}

\begin{figure}
\centering           
\includegraphics[totalheight=2.5in]{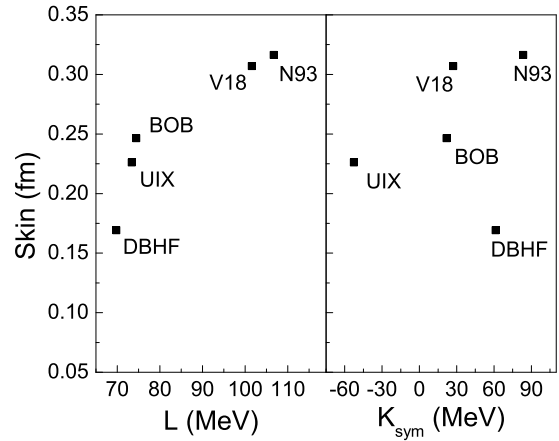} 
\vspace*{-0.3cm}
\caption{Left panel: neutron skin of $^{208}$Pb {\it vs.} the symmetry pressure for the models
considered in the text; Right panel: 
neutron skin of $^{208}$Pb {\it vs.} the curvature of the symmetry energy, $K_{sym}$, for the same models. 
} 
\label{SLK}
\end{figure}

In Fig.~\ref{esmicro}, we display Idaho DBHF predictions for the symmetry energy, solid black curve, 
along with those from V18, BOB, UIX, and N93. 
All values of the symmetry energy at the respective saturation densities are 
between 29 and 34 MeV. A larger spreading is seen in the $L$ parameter, 
see left panel of Fig.~\ref{SLK}, where the values range from about 70 to 106 MeV. The respective neutron skin predictions are shown on the vertical axis.

We notice that all BHF+TBF models predict larger $L$, and thus larger neutron skins, as compared with DBHF, corresponding to 
a faster growth of the energy per particle in neutron matter relative to symmetric matter.               
This can be seen from Fig.~\ref{esmicro}, especially for the higher
densities.                                                                          
The present calculations reveal that there are more subtle, but significant differences at low to medium densities as well.

Some comments are in place concerning 
the nature of the $L$ {\it vs.} $S$ correlation.                  
If a family 
of models is constructed so that they differ {\it mostly} in the slope of the symmetry energy, a linear 
correlation may be expected among $L$ and the skin.\cite{Li05} The models we are considering, however, 
differ from  one another in more than just the slope of the symmetry energy. Thus, although the general 
pattern is that a larger $L$ corresponds to a larger skin, the relation is more complex, as the r.m.s.
radius of the neutron distribution will, to some extent, receive feedback from the smaller or larger
degree of attraction that binds neutrons and protons, and which depends on the symmetric matter EoS. 

The next term in the expansion of the symmetry energy is the 
$K_{sym}$ parameter,
\begin{equation}
K_{sym} = 9 \rho_0^2 \Big (\frac{\partial^2 e_{sym}(\rho)}{\partial \rho ^2}\Big )_{\rho_0} \, , 
\label{Ksym}
\end{equation} 
which is a measure for the curvature of the symmetry energy. The neutron skin {\it vs.\ } $K_{sym}$ 
is shown in the right panel of Fig.~\ref{SLK} for the various models. Although the  
values of $K_{sym}$ appear more spread out, the large negative values obtained with 
some of the parametrizations of the Skyrme model are not present.     
Those                        
large negative values (as low as -600 MeV) produced by Skyrme models indicate a strongly downward
curvature of the symmetry energy already at low to medium densities.                                          
We also notice from Fig.~\ref{SLK} (right panel) 
that no clear correlation can be identified between $K_{sym}$ and the neutron skin.

\begin{figure}[!t]
\centering          
\includegraphics[totalheight=2.5in]{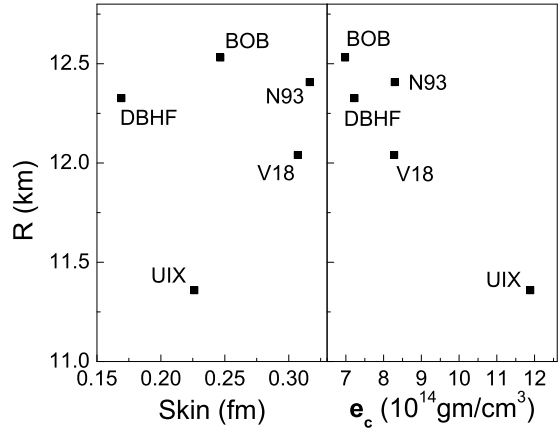}  
\vspace*{-0.2cm}
\caption{Left panel: neutron skin of $^{208}$Pb {\it vs.} the radius of a 1.4M$_{\odot}$ neutron star for the models
considered in the text. Right panel: star radius {\it vs.} the central energy density for the same models. 
} 
\label{RS}
\end{figure}

We conclude this section by showing in Fig.~\ref{RS} the relation between the neutron skin of $^{208}$Pb and the radius  
of a 1.4M$_{\odot}$ neutron star for the models we are considering. (A more detailed discussion of neutron stars and the EoS will be presented in Sec.~4.) Stellar matter contains neutrons in $\beta$
equilibrium with protons, electrons, and muons. 
Tabulated values for the latest Idaho DBHF EoS can be found later in this article. We have applied $\beta$-stability in the same way to all the 
various models of        
 Ref.\cite{Catania4} starting from the given parametrized versions of the respective symmetric matter and neutron matter EoS.
For the case of leptons, typically treated as a gas of non-interacting fermions, the equations of $\beta$-stability
(which amount to imposing energy conservation and charge neutrality), 
are elementary to solve (see Sec.~4.2). 

At subnuclear densities, all the EoS considered here are joined with the crustal equations of state from  the work of Harrison and Wheeler\cite{HW} (for energy densities between 10 and 10$^{11}$ g~cm$^{-3}$) and the work of Negele and 
Vautherin\cite{NV} (for energy densities less than 1.7$\times$10$^{13}$g~cm$^{-3}$). 

At first, Fig.~\ref{RS} can appear surprising, since it shows that 
 larger skin does not necessarily imply larger radius. On the other hand, one must keep in mind that, 
 as we argued previously, these 
models differ from  one another in more than just the slope of the symmetry energy. Furthermore, the star 
radius probes higher densities than the skin does. 
Generally, if a family 
of models is constructed so that they differ {\it mostly} in the slope of the symmetry energy, a linear 
correlation is expected among $L$, the skin, and $R$.\cite{Moust07}

In summary:                     
Most models agree on the value of the symmetry energy around the saturation point within a few MeV, 
but we are far from a reasonable agreement on its derivative                   
 and even farther from agreement on its curvature.
Although microscopic models do not display as much spread as phenomenological ones, 
a  point that comes out clearly from the present study is that 
a measurement of the neutron skin of $^{208}$Pb with an accuracy of 0.05 fm, as it has been announced,\cite{Hor} should definitely
be able to discriminate among the EoS from these microscopic models.

\subsubsection{Neutron and proton single-particle properties and the symmetry potential}

In this section we will concentrate specifically on predictions of           
isovector single-particle properties, such as the neutron and proton single-particle 
potentials and the closely related symmetry potential. These ``observables"           
depend 
sensitively on the difference between neutron and proton properties in asymmetric
matter and play an important role in simulations of heavy-ion collisions
with neutron-rich nuclei.

In terms of the $G$-matrix, calculated as described in Sec.~3.2.1,               
we write the single-nucleon potential                            
(in the case of unequal Fermi levels for protons 
and neutrons), as                                        
\begin{equation}
U_i(k) = Re[\sum_{q<k_F^n} <kq|G_{in}|kq-qk>
+ \sum_{q<k_F^p} <kq|G_{ip}|kq-qk>] \, , 
\label{Uik} 
\end{equation}
where $i$ = $n/p$ for neutron/proton, and 
 $k$  refers to states below and above the Fermi momentum (consistent with the ``continuous choice" for the 
single-particle spectrum). 

\begin{figure}[!t]
\centering          
\vspace*{-1.0cm}  
\includegraphics[totalheight=3.5in]{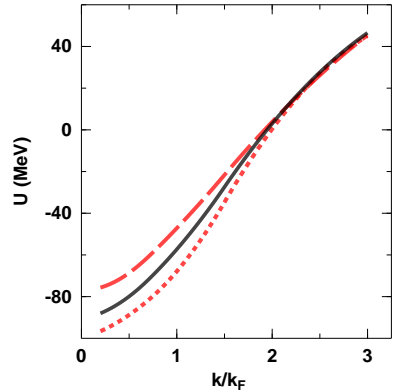}  
\vspace*{-2.2cm}
\caption{ The neutron (dashed line) and proton (dotted) single-particle potentials as a function of the 
momentum at fixed average density and  
$\alpha$=0.4. The solid curve in the middle is the single-nucleon potential at $\alpha$=0. 
} 
\label{unp}
\end{figure}

We begin by examining the momentum dependence of 
$U_{n/p}$, the single neutron/proton potential in neutron-rich matter. 
In Fig.~\ref{unp}, we show 
$U_{n/p}$ as a 
function of the momentum and a fixed value of the asymmetry parameter,
$\alpha=$ 0.4. The middle curve shows $U_n$=$U_p$ at $\alpha=$0.                 
The total nucleon density considered in the figure is equal to 0.185 fm$^{-3}$ and
corresponds to a Fermi momentum of 1.4 fm$^{-1}$, which is very close to our predicted
saturation density.

 For increasing values of 
$\alpha$ (obviously, $U_n=U_p$ for $\alpha$=0), the proton potential becomes more attractive while the opposite
tendency is observed in $U_n$. This reflects the fact that the proton-neutron interaction, the 
one predominantly felt by the single proton as the proton density is depleted, is more
attractive than the one between identical nucleons. 
Also, as it appears reasonable, the dependence on $\alpha$ becomes weaker at larger momenta.

The role of the momentum dependence of the symmetry potential in heavy-ion collisions was            
examined and found to be important. Symmetry potentials with and without 
momentum dependence can lead to significantly
different predictions of collision observables.\cite{DAS04} More recently, these issues have been 
revisited in hot asymmetric matter.\cite{Moust207,Moust08}

\begin{figure}[!t]
\centering          
\vspace*{1.0cm}  
\includegraphics[totalheight=3.0in]{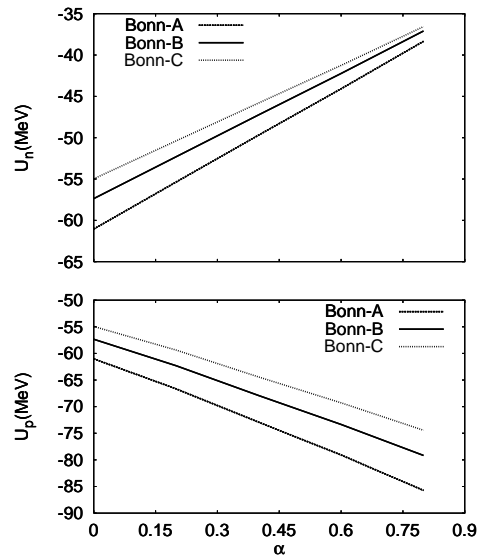}  
\vspace*{0.2cm}
\caption{ The neutron and proton single-particle potentials as a function of the 
asymmetry parameter at fixed average density and momentum equal to the average 
Fermi momentum. 
} 
\label{ualpha}
\end{figure}

Regarding                                                                      
$U_{n/p}$ as functions of                                    
the asymmetry parameter $\alpha$, one can easily
verify that the following approximate relation applies:                
\begin{equation}
U_{n/p}(k,k_F,\alpha) \approx U_{n/p}(k,k_F,\alpha=0) \pm U_{sym}(k,k_F)\alpha \, , 
\label{Unp}
\end{equation}
with the $\pm$ referring to neutron/proton, respectively.                   
Figure~\ref{ualpha}  displays the left-hand side of Eq.~(\ref{Unp}) for fixed density and nucleon momentum and
clearly reveals the linear 
behaviour of                                   
$U_{n/p}$ as a function of $\alpha$. Thus,                 
one can expect isospin splitting of the single-particle potentials to be effective in separating           
the collision dynamics of neutrons and protons. 
Although the main focus of Fig.~\ref{ualpha} is the $\alpha$ dependence, 
predictions are displayed for three different potentials, Bonn A,\cite{Mac89} B,\cite{Mac89} and C.\cite{Mac89} 
These three models differ mainly in the strength of the tensor force, which
is mostly carried by partial waves with isospin equal to 0 and thus should fade away
in the single-neutron potential
as the neutron fraction increases. In fact, the figure demonstrates reduced differences among the values               
of $U_n$ predicted with the three potentials at large $\alpha$.

Already several decades ago, it was pointed out that the real part of the nuclear 
optical potential depends on the asymmetry parameter as in Eq.~(\ref{Unp}).\cite{Lane}  
Then,                                                                   
the quantity 
\begin{equation}
\frac{U_{n} + U_p}{2} = U_0 ,               
\label{U0}
\end{equation}
which is obviously the single-nucleon potential in absence of asymmetry,
should be a reasonable approximation to the isoscalar part of the optical 
potential in the interior of a nucleus. The momentum dependence of $U_0$ (which is shown in Fig.~\ref{unp} as the 
$\alpha$=0 curve), is important for extracting information about the symmetric matter EoS
and is reasonably agreed upon
.\cite{u02,u03,u04,u05,u06,u07,u08,u09} 

\begin{figure}[!t]
\centering          
\vspace*{-2.5cm}
\includegraphics[totalheight=5.5in]{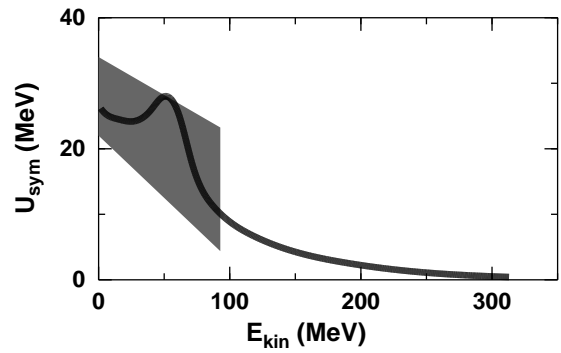}  
\vspace*{-4.5cm}
\caption{ The symmetry potential close to saturation density.                                              
} 
\label{usym}
\end{figure}

On the other hand, 
\begin{equation}
\frac{U_{n} - U_p}{2\alpha} = U_{sym}             
\label{Usym} 
\end{equation}
should be comparable with the Lane potential,\cite{Lane} or the isovector 
part of the nuclear optical potential. (In the two equations
above the dependence upon density, momentum, and asymmetry has been suppressed for
simplicity.) 
We have calculated $U_{sym}$ as a function of
the momentum, or rather the corresponding kinetic energy. 
The predictions obtained with Bonn B are shown in Fig.~\ref{usym}.
They are compared with the phenomenological expression\cite{Lane}
\begin{equation}
U_{Lane}      = a -b T \, ,                       
\label{Lane}
\end{equation}
where $T$ is the kinetic energy, $a \approx 22-34 MeV$, $b\approx 0.1-0.2 MeV$.

We observe that the strength of the (DBHF) predicted symmetry potential decreases
with energy, a behavior which is consistent with the empirical information.   
The same comparison is done in 
  Ref.\cite{BAL04} starting from a                             
  phenomenological formalism for the single-nucleon potential.\cite{Bomb01,Rizzo}  
There, it is shown that it is 
 possible to choose two sets of parameters which lead to similar values of the 
 symmetry energy but exactly opposite tendencies in the energy dependence of the symmetry
 potential as well as 
 opposite sign of the proton-neutron mass splitting.
 As a consequence of that, these two sets of parameters lead to very different
 predictions for observables in heavy-ion collisions induced by neutron-rich nuclei.\cite{Rizzo}
 This fact suggests that                                                                 
constraints from ``differential" observables, namely those  specifically sensitive to the difference
between proton and neutron properties in asymmetric matter, are very much needed.

\begin{figure}[!t]
\centering          
\vspace*{3.0cm}  
\includegraphics[totalheight=2.0in]{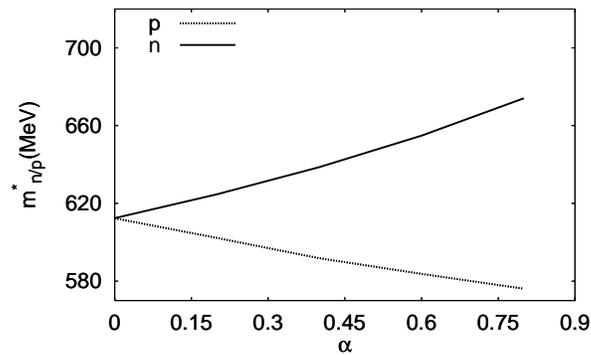}  
\vspace*{-2.0cm}
\caption{ The neutron and proton effective mass  as a function of the             
asymmetry parameter at fixed average density.                                   
} 
\label{mnp}
\end{figure}

The effective masses for proton and neutron corresponding to the single-nucleon potentials 
of Fig.~\ref{unp}
 are shown in Fig.~\ref{mnp} as a function
of $\alpha$.                                                                    
 The predicted effective mass of the neutron being larger than the 
proton's is a trend shared with 
microscopic non-relativistic
calculations.\cite{BL91} In the non-relativistic case, one can show from elementary arguments
based on the curvature of the (parabolic) single-particle potential that a more attractive 
potential, as the one of the proton, leads to a smaller effective mass.
In the DBHF  quasi-particle approximation described in Sec.~3.2.1, one 
assumes momentum-independent nucleon self-energies, $U_S$ and $U_V$, with a vanishing
spacial component of the vector part. In such limit, the ``Dirac mass" and the non-relativistic or ``Landau mass" 
should coincide to leading order in the expansion of the relativistic single-particle potential.\cite{Fuchs}
It is therefore not surprising that 
our proton and neutron effective masses, although defined as Dirac masses ($m^*_i = m+U_{S,i}$), display
a trend that is similar to the one generally expected in the non-relativistic case. Most likely, the main reason for the 
discrepancy in the predicted isospin mass splitting between the Idaho DBHF calculation and the DBHF model from Ref.\cite{Fuchs} 
is in the way the scalar and vector fields are treated, particularly their momentum dependence. 

Closely related to the effective mass is the concept of in-medium effective cross section. 
Transport equations, such as the Boltzmann-Uehling-Uhlenbeck equation, 
describe the evolution of a non-equilibrium gas of 
strongly interacting hadrons drifting                
in the presence of the mean field while undergoing    
two-body collisions. Thus heavy ion collision simulations require the                                      
knowledge of in-medium two-body cross sections as well as the mean field.
Both should be calculated microscopically and                 
self-consistently starting from the basic nuclear forces\cite{sig2}
and are closely related to the EoS.

\begin{table}[pt]
\tbl                                                                            
{$np$ total effective cross sections in symmetric matter calculated with 
various many-body models, see text for details.                            
}       
{\begin{tabular}{@{}ccccc@{}} \toprule
$k_{F}(fm^{-1})$ & $q_0(MeV)$ & $\sigma _{np}^{DBHF}(mb)$ &
                   $\sigma _{np}^{BHF+TBF}(mb)$ & $\sigma _{np}^{BHF}(mb)$    
                                            \\     \colrule                 
1.1 &   250  &  34.38& 44.65 &51.74         \\                  
1.1 &   300  &  23.14& 29.01 &31.85           \\                
1.1 &   350  &  20.63& 23.56 &25.62           \\                
1.4 &   250  &  26.74& 31.25 &39.82           \\                
1.4 &   300  &  17.26& 25.28 &30.96           \\                
1.4 &   350  &  16.77& 21.17 &25.23           \\                
1.7 &   250  &  17.20& 19.03 &29.12           \\                
1.7 &   300  &  15.06& 17.59 &25.02           \\                
1.7 &   350  &  12.33& 13.99 &21.04           \\                
\botrule 
\end{tabular}}
\end{table}
                                                                       
\begin{table}[pt]
\tbl                                                                            
{As in the previous Table, for identical particles.}                        
{\begin{tabular}{@{}ccccc@{}} \toprule
$k_{F}(fm^{-1})$ & $q_0(MeV)$ & $\sigma _{NN}^{DBHF}(mb)$ &
                   $\sigma _{NN}^{BHF+TBF}(mb)$ & $\sigma _{NN}^{BHF}(mb)$  \\ 
                                                   \colrule                 
1.1 &   250  &  18.00& 18.15 &22.98         \\                  
1.1 &   300  &  16.41& 17.47 &19.48           \\                
1.1 &   350  &  17.08& 18.55 &19.51           \\                
1.4 &   250  &  15.72& 13.74 &19.76           \\                
1.4 &   300  &  13.70& 14.43 &17.23           \\                
1.4 &   350  &  16.31& 16.89 &19.46           \\                
1.7 &   250  &  18.05&  7.87 &13.36           \\                
1.7 &   300  &  17.93&  9.98 &13.98          \\                
1.7 &   350  &  13.96& 11.48 &16.43           \\                
\botrule 
\end{tabular}}
\end{table}

To be fully applicable throughout the evolution of the colliding system, where high densities may be involved, 
in-medium cross sections must go beyond the conventional BHF. 
Although an extensive discussion of in-medium cross sections will not be presented here,
we show in Tables 1 and 2 a comparison of the latest Idaho (DBHF) in-medium cross sections and those 
obtained from the BHF+TBF model of Lombardo {\it et al.}.\cite{sig2}

We recall that the relativistic elastic differential cross section is calculated from 
\begin{equation}
 \sigma(\theta) = \frac{(m^*)^4}{4 \pi ^2 \hbar ^4 s^*} |G(\theta)|^2 \, , 
\label{sigrel}
\end{equation} 
where $s^* = 4((m^*)^2 + {\vec q}^2)$. 
Notice that the relativistic amplitude, $G$, is related to the non-relativistic one through 
$G = \frac{E^*}{m^*}G_{n.r.}$, with           
 $E^* = ((m^*)^2 + {\vec q}^2)^{1/2}$. 

In Tables 1 and 2, $q_0$ signifies the momentum of either nucleon in their center-of-mass frame, for simplicity assumed to coincide with 
the nuclear matter rest frame. The 4th and 5th columns show the results obtained by Lombardo {\it et al.} 
with BHF calculations implemented or not with TBF. There is generally a fair amount of agreement between 
the DBHF and the BHF+TBF results. For 
the $np$ case, energy and density dependence appear consistent among the two sets of results displayed 
in columns 3 and 4, although the BHF+TBF results are generally larger. For the case of identical nucleons,
the DBHF values are in good agreement with those from BHF+TBF, with the exception of the 
highest densities. 
This is to be expected in view of the larger differences existing between the DBHF and the BHF+TBF predictions of the EoS at high density.

Besides being a crucial part of the input for transport models, in-medium effective
cross sections are important in their own right as they allow to establish an immediate connection
with the nucleon mean free path, $\lambda$, one of the most fundamental properties characterizing the 
propagation of nucleons through matter.\cite{mfp} It can be written as 
\begin{equation}
\lambda = \frac{1}{\sigma _{pp} \rho _p + \sigma _{pn} \rho _n} \, , 
\label{lambda} 
\end{equation}
for a proton propagating through matter with proton and neutron densities equal to $\rho _p$ and 
$\rho _n$, respectively. 
The mean free path enters the calculation of the nuclear 
transparency function, which is the probability that the projectile will pass through the target without interacting.
This is closely related to the total reaction 
cross section of a nucleus, which can be used to extract
 nuclear r.m.s. radii within Glauber models.\cite{Glauber} Therefore,  microscopic in-medium 
 isospin-dependent NN cross sections can ultimately help obtain 
information about the properties of exotic, neutron-rich nuclei, such as, for instance, the target density.
The latter is a very useful information, especially           
for exotic nuclei or nuclei with large neutron skins (where densities cannot be probed with conventional
charged lepton scattering). 

\section{Neutron Star Properties}
Proceeding in  our review of the various systems and phenomena where 
the EoS plays a chief role, we will now take a look at 
some of the most exotic systems in the universe, compact stars.                        

\subsection{A brief review of neutron star structure and available constraints} 

Neutron stars are stable configurations containing the most dense form of matter found in the universe.
They are therefore unique laboratories to study the properties of highly compressed (cold) matter. 
Furthermore, the possibility of exploring the structure of neutron stars via gravitational waves\cite{GW} 
makes these exotic objects even more exciting. 

The densities found in neutron stars range from the density of iron to several times normal nuclear density. 
In the low-density region, matter is highly compressed and fully ionized and consists of electrons and ions of iron.
As density increases, 
charge neutrality requires matter to become more neutron rich. In this density range
($10^7 \leq \rho \leq 10^{11}$~g~cm$^{-3}$), neutron-rich nuclei appear, such as isotopes of
$^{62}$Ni,
$^{86}$Kr,
$^{84}$Se,
$^{80}$Zn,
$^{124}$Mo,
$^{122}$Zr.
Above densities of approximately $ 10^{11}$~g~cm$^{-3}$, free neutrons begin to form a continuum of states.
The inner crust is then a compressed solid with a fluid of neutrons. 
At densities equal to approximately 1/2 of saturation density, clusters begin to merge into a continuum. In this phase, matter is a uniform fluid 
of neutrons, protons, and leptons. Protons and neutrons can condense in superfluids and superconducting states 
below temperatures of about 10$^9$~K.\cite{Sedr06} 
Above a few times nuclear matter density, the actual composition of stellar matter is not known.  
Strange baryons can appear when the nucleon chemical potential is of the order of their rest mass. 
We will address the issue of strange matter later in this article. 
At even higher densities, transitions to other phases are speculated, such as a deconfined, rather than hadronic, phase. The critical density for
such transition cannot be predicted reliably because it lies in a range where QCD in non perturbative.\cite{Sedr06}

Recently, a semi-algebraic expression for the EoS of cold quark matter has been derived within the 
Dyson-Schwinger formalism.\cite{Klaen09} 
The Dyson-Schwinger framework can address both confinement and dynamical chiral symmetry breaking,\cite{Klaen09} 
unlike other models commonly used in conjunction with cold quark matter, such as the Nambu-Jona-Lasinio model or
bag-type models. It will be interesting to see applications of the model described in Ref.\cite{Klaen09} 
to dense astrophysical systems. For that purpose, the quark matter EoS will need to be supplemented by
the nuclear matter EoS. In the meantime, it is encouraging that the model predicts coincident 
deconfinement and chiral symmetry restoring transitions.

The possibility has been speculated that the most stable state at zero pressure may be
$u$, $d$, $s$ quark matter instead of iron. This would imply that strange quark matter is the most
stable (in fact, the absolutely stable) state of strongly interacting matter, as originally proposed 
by Bodmer,\cite{Bodmer} Witten,\cite{Witten} and Terazawa.\cite{Teraz}
In such case, hyperonic and hybrid stars would have to be metastable with respect to stars composed 
of stable three-flavor strange quark matter,\cite{Weber} which is lower in energy 
than two-flavor quark matter due to the extra Fermi levels open to strange quarks. 
Whether or not strange quark stars can give rise to pulsar glitches, (which are observed sudden small changes 
in the rotational frequency of a pulsar),
may be a decisive test of the strange quark matter hypothesis.\cite{Weber}

The maximum mass and the radius of a neutron star are sensitive to different aspects of the EoS. The maximum mass
is mostly determined by the stiffness of the EoS at densities greater than a few times saturation density.
Non-nucleonic degrees of freedom, which typically make their appearance at those densities, are
known to have a considerable impact on the maximum mass of the star. 
Causality imposes a limit of about 3 solar masses to the maximum mass, the existence of which is inherent to general relativity
and is predicted by the equation of hydrostatic equilibrium,
\begin{equation}
\frac{d P(r)}{dr} = -\frac{G}{c^2}\frac{(P(r)+\epsilon(r))(M(r)+4\pi r^3 P(r)/c^2)}{r(r-2GM(r)/c^2)}  \; , 
\label{GR1}
\end{equation}
with 
\begin{equation}
\frac{d M(r)}{dr} = 4\pi ^2 \rho(r) \, ,                           
\label{GR2}
\end{equation}
where $\epsilon$ is the total mass-energy density. 
The pressure is related to the energy/particle through 
\begin{equation}
P(\rho) = \rho ^2 \frac{\partial e(\rho)}{\partial \rho} \, .                                      
\label{Pr}
\end{equation}
It's worth recalling that no mass limit exists in Newtonian gravitation.

The star radius is mainly sensitive to the slope of the symmetry energy. In particular, it is closely connected 
to the  internal pressure (that is, the energy gradient) of matter at
densities between about 1.5$\rho_0$ and 
2-3$\rho_0$.\cite{Latt07} 

Lattimer and Prakash investigated the maximum central density question.\cite{Latt07}
Combining the causality limit ($R\geq$2.87$GM/c^2$) with the central density-mass relation implied by the 
Tolman VII solution,\cite{Tolman} 
\begin{equation}
\rho= \rho _{c} (1 - r^2/R^2) \, ,                                 
\label{Tol}
\end{equation}
they obtained the relation                                                    
\begin{equation}
\rho _{c} M^2 \leq 15.3 \times 10^{15} M_{\odot}^2 \, g\,cm^{-3} \, . 
\label{ecmax}
\end{equation}

The maximum gravitational mass and the corresponding radius are the typical observables 
used to constraint the EoS.                                                                          
Much more stringent constraints could be imposed on EoS if these two quantities could be 
determined independently from each other. At this time, the heaviest neutron star (with        
accurately known mass) has a mass of 1.671 $\pm$ 0.008 $M_{\odot}$,\cite{Champ08} but neutron stars
with masses above 2 solar masses are expected to exist.\cite{Fre09}  
Mass limits are obtained from observations of binary systems, either a two-neutron star system 
or a neutron star and a massive companion, such as a white dwarf. The pulsar in the Hulse-Taylor
binary system\cite{Fre09} has a mass of 
1.4408$\pm$ 0.0003 $M_{\odot}$, to date the best mass determination.

Measurements of the radius are considerably less precise than mass measurements.\cite{Latt07}
The radiation or photospheric radius, $R_{\infty}$, is related to the actual stellar radius by 
$R_{\infty} = R(1-R/R_S)^{-1/2}$, where $R_S=2GM/R$ is the Schwarzschild radius. 
Estimates are usually based on                                              
thermal emission of cooling stars, including redshifts, 
and the properties of sources with bursts or thermonuclear explosions at the surface. 

Neutrinos are also used as a probe of the EoS. 
Neutrino luminosity is controlled by several factors including the total mass of the (proton-neutron)
star and the opacity of neutrinos at high densities, which of course is sensitive to the 
EoS of dense matter.

Gravitational waves\cite{GW} are a less conventional way to probe neutron star properties. 
Emission of gravitational waves causes orbital decay, the observation of which would allow 
an estimate of the moment of inertia and thus, together with an accurate measurement of the 
mass, impose stronger boundaries on the EoS. 
A measurement of the moment of inertia within 10\%, together with the information on the mass, would 
be able to discriminate among various EoS.\cite{Latt07}
To date, the best determination of the moment of inertia is the one for the Crab pulsar\cite{Crab}
which would rule out only very soft EoS.\cite{Latt07}

The minumum mass of a neutron star is also a parameter of interest. For a cold, stable system, 
the minimum mass is about 
0.09 $M_{\odot}$.\cite{Latt07}                                        
The smallest reliably estimated neutron star mass is the companion of the binary pulsar 
J1756-2251, which has a mass of 
1.18$\pm$ 0.02 $M_{\odot}$.\cite{Faulk}

\subsection{Comparing predictions of microscopic models}

As constraints from nuclear physics and/or astrophysics  promise to become more stringent, it is important to understand and compare how the nature of the various predictions
is related to the features of each model. In microscopic approaches, the 
tight connection with the underlying forces facilitates physical interpretation of the predictions 
in terms of the characteristics of the nuclear force and its behavior in the medium. 
Motivated by these considerations, in this section  
we calculate several neutron star properties, for static and/or rotating stars, using 
equations of state based on different microscopic models.                       
These will be the same as those used in Sec.~3.2.3 
to examine symmetry energy and neutron skin predictions, namely BOB, V18, N93, and UIX, 
along with Idaho DBHF. 

In Sec.~3.2.3,                                                     
we compared the 
predictions of the neutron skin in $^{208}$Pb by these models and correlated them  
with differences in the slope of the symmetry energy. 
Model differences become larger at high-density and will naturally impact neutron star predictions. 

\begin{table}[pt]
\tbl{DBHF equation of state of pure neutron matter.} 
{\begin{tabular}{@{}ccc@{}} \toprule
Baryon density(1/$cm^3$)  & Energy density($g/cm^3$)  & Pressure($dyne/cm^2$)  \\                                         
\colrule
 0.145902E+38  &  0.245257E+14  &  0.434992E+32 \\
 0.231688E+38  &  0.389881E+14  &  0.874118E+32 \\
 0.345843E+38  &  0.582634E+14  &  0.156362E+33 \\
 0.492421E+38  &  0.830558E+14  &  0.257664E+33 \\
 0.675475E+38  &  0.114075E+15  &  0.468635E+33 \\
 0.781946E+38  &  0.132151E+15  &  0.621198E+33 \\
 0.899057E+38  &  0.152061E+15  &  0.816935E+33 \\
 0.102731E+39  &  0.173905E+15  &  0.111487E+34 \\
 0.116722E+39  &  0.197791E+15  &  0.156721E+34 \\
 0.131929E+39  &  0.223833E+15  &  0.223210E+34 \\
 0.148402E+39  &  0.252158E+15  &  0.319807E+34 \\
 0.166192E+39  &  0.282905E+15  &  0.463315E+34 \\
 0.185350E+39  &  0.316247E+15  &  0.681545E+34 \\
 0.205927E+39  &  0.352381E+15  &  0.993009E+34 \\
 0.227973E+39  &  0.391534E+15  &  0.141505E+35 \\
 0.251538E+39  &  0.433969E+15  &  0.203050E+35 \\
 0.276674E+39  &  0.480078E+15  &  0.293314E+35 \\
 0.303432E+39  &  0.530247E+15  &  0.409031E+35 \\
 0.331861E+39  &  0.584918E+15  &  0.558832E+35 \\
 0.362012E+39  &  0.644589E+15  &  0.746722E+35 \\
 0.393937E+39  &  0.709762E+15  &  0.965629E+35 \\
 0.427685E+39  &  0.780907E+15  &  0.122694E+36 \\
 0.463308E+39  &  0.858621E+15  &  0.152765E+36 \\
 0.500856E+39  &  0.943372E+15  &  0.185795E+36 \\
 0.540380E+39  &  0.103570E+16  &  0.223317E+36 \\
 0.581930E+39  &  0.113630E+16  &  0.268106E+36 \\
 0.625557E+39  &  0.124615E+16  &  0.324163E+36 \\
 0.671312E+39  &  0.136650E+16  &  0.395072E+36 \\
 0.719245E+39  &  0.149890E+16  &  0.485417E+36 \\
 0.769408E+39  &  0.164502E+16  &  0.589746E+36 \\
 0.821850E+39  &  0.180610E+16  &  0.707244E+36 \\
 0.876622E+39  &  0.198388E+16  &  0.846053E+36 \\
 0.933776E+39  &  0.218030E+16  &  0.100924E+37 \\
 0.993361E+39  &  0.239765E+16  &  0.120127E+37 \\
 0.105543E+40  &  0.263842E+16  &  0.142307E+37 \\
 0.112003E+40  &  0.290521E+16  &  0.167838E+37 \\
 0.118721E+40  &  0.320102E+16  &  0.197388E+37 \\
 0.125703E+40  &  0.352921E+16  &  0.231442E+37 \\
 0.132954E+40  &  0.389345E+16  &  0.270588E+37 \\
 0.140478E+40  &  0.429785E+16  &  0.315475E+37 \\
 0.148280E+40  &  0.474691E+16  &  0.366826E+37 \\
 0.156366E+40  &  0.524560E+16  &  0.425495E+37 \\
 0.164741E+40  &  0.579940E+16  &  0.492059E+37 \\
 0.173410E+40  &  0.641418E+16  &  0.567978E+37 \\
 0.182378E+40  &  0.709679E+16  &  0.654219E+37 \\
\botrule
\end{tabular}}
\end{table}

\begin{table}[pt]
\tbl{As in the previous Table, for $\beta$-equilibrated matter.}                                       
{\begin{tabular}{@{}ccc@{}} \toprule
Baryon density(1/$cm^3$)  & Energy density($g/cm^3$)  & Pressure($dyne/cm^2$)  \\                                         
\colrule
 0.145902E+38  &  0.245236E+14  &  0.416322E+32 \\
 0.231688E+38  &  0.389827E+14  &  0.822804E+32 \\
 0.345843E+38  &  0.582501E+14  &  0.142104E+33 \\
 0.492421E+38  &  0.830269E+14  &  0.230061E+33 \\
 0.675475E+38  &  0.114019E+15  &  0.409814E+33 \\
 0.781946E+38  &  0.132073E+15  &  0.540126E+33 \\
 0.899057E+38  &  0.151955E+15  &  0.711724E+33 \\
 0.102731E+39  &  0.173765E+15  &  0.970498E+33 \\
 0.116722E+39  &  0.197605E+15  &  0.137046E+34 \\
 0.131929E+39  &  0.223589E+15  &  0.201627E+34 \\
 0.148402E+39  &  0.251837E+15  &  0.261928E+34 \\
 0.166192E+39  &  0.282481E+15  &  0.330297E+34 \\
 0.185350E+39  &  0.315683E+15  &  0.517205E+34 \\
 0.205927E+39  &  0.351619E+15  &  0.757028E+34 \\
 0.227973E+39  &  0.390495E+15  &  0.110037E+35 \\
 0.251538E+39  &  0.432555E+15  &  0.160843E+35 \\
 0.276674E+39  &  0.478138E+15  &  0.235092E+35 \\
 0.303432E+39  &  0.527601E+15  &  0.334557E+35 \\
 0.331861E+39  &  0.581355E+15  &  0.465834E+35 \\
 0.362012E+39  &  0.639861E+15  &  0.632143E+35 \\
 0.393937E+39  &  0.703594E+15  &  0.832228E+35 \\
 0.427685E+39  &  0.773030E+15  &  0.107139E+36 \\
 0.463308E+39  &  0.848703E+15  &  0.134775E+36 \\
 0.500856E+39  &  0.931088E+15  &  0.165606E+36 \\
 0.540380E+39  &  0.102069E+16  &  0.200271E+36 \\
 0.581930E+39  &  0.111814E+16  &  0.241417E+36 \\
 0.625557E+39  &  0.122434E+16  &  0.293125E+36 \\
 0.671312E+39  &  0.134047E+16  &  0.358691E+36 \\
 0.719245E+39  &  0.146797E+16  &  0.442173E+36 \\
 0.769408E+39  &  0.160841E+16  &  0.539098E+36 \\
 0.821850E+39  &  0.176298E+16  &  0.649283E+36 \\
 0.876622E+39  &  0.193337E+16  &  0.779667E+36 \\
 0.933776E+39  &  0.212141E+16  &  0.932987E+36 \\
 0.993361E+39  &  0.232930E+16  &  0.111395E+37 \\
 0.105543E+40  &  0.255941E+16  &  0.132452E+37 \\
 0.112003E+40  &  0.281437E+16  &  0.156916E+37 \\
 0.118721E+40  &  0.309721E+16  &  0.185697E+37 \\
 0.125703E+40  &  0.341144E+16  &  0.219046E+37 \\
 0.132954E+40  &  0.376060E+16  &  0.257227E+37 \\
 0.140478E+40  &  0.414881E+16  &  0.301377E+37 \\
 0.148280E+40  &  0.458067E+16  &  0.352202E+37 \\
 0.156366E+40  &  0.506130E+16  &  0.410655E+37 \\
 0.164741E+40  &  0.559636E+16  &  0.477543E+37 \\
 0.173410E+40  &  0.619202E+16  &  0.554149E+37 \\
 0.182378E+40  &  0.685522E+16  &  0.641233E+37 \\
\botrule
\end{tabular}}
\end{table}

\begin{table}[pt]
\tbl{Equation of state used for the crust. See text for details.}                                       
{\begin{tabular}{@{}ccc@{}} \toprule
Baryon density(1/$cm^3$)  & Energy density($g/cm^3$)  & Pressure($dyne/cm^2$)  \\                                         
\colrule
  0.59701000E+25 &  0.99998600E+01 &  0.40721410E+12\\
  0.19099900E+26 &  0.31992190E+02 &  0.47174340E+14\\
  0.38376300E+26 &  0.64279760E+02 &  0.33550870E+15\\
  0.15492600E+27 &  0.25949850E+03 &  0.88776630E+16\\
  0.31128200E+27 &  0.52139300E+03 &  0.38114990E+17\\
  0.62543700E+27 &  0.10476010E+04 &  0.15215920E+18\\
  0.25249000E+28 &  0.42291870E+04 &  0.20992040E+19\\
  0.50731100E+28 &  0.84974150E+04 &  0.74261220E+19\\
  0.20480200E+29 &  0.34304300E+05 &  0.87378890E+20\\
  0.41149300E+29 &  0.68925200E+05 &  0.29327530E+21\\
  0.13164600E+30 &  0.22050990E+06 &  0.21592540E+22\\
  0.26450500E+30 &  0.44305570E+06 &  0.70850350E+22\\
  0.53144400E+30 &  0.89020260E+06 &  0.23123900E+23\\
  0.21453100E+31 &  0.35937570E+07 &  0.19523170E+24\\
  0.43102100E+31 &  0.72207080E+07 &  0.54232090E+24\\
  0.13787700E+32 &  0.23100980E+08 &  0.28166750E+25\\
  0.34951600E+32 &  0.58569360E+08 &  0.10057550E+26\\
  0.70215700E+32 &  0.11767960E+09 &  0.25560490E+26\\
  0.28334500E+33 &  0.47507410E+09 &  0.15831900E+27\\
  0.56915000E+33 &  0.95453350E+09 &  0.38815460E+27\\
  0.22959600E+34 &  0.38534730E+10 &  0.22901680E+28\\
  0.46107400E+34 &  0.77425110E+10 &  0.55287600E+28\\
  0.14734400E+35 &  0.24770450E+11 &  0.23888140E+29\\
  0.29580200E+35 &  0.49769610E+11 &  0.57354730E+29\\
  0.59374600E+35 &  0.99998600E+11 &  0.13755860E+30\\
  0.30685800E+36 &  0.51813430E+12 &  0.66738040E+30\\
  0.60373600E+36 &  0.10203670E+13 &  0.12773240E+31\\
  0.10985200E+37 &  0.18580330E+13 &  0.20087290E+31\\
  0.15931600E+37 &  0.26960450E+13 &  0.26615840E+31\\
  0.19888300E+37 &  0.33666280E+13 &  0.32106350E+31\\
  0.23844100E+37 &  0.40373370E+13 &  0.38075420E+31\\
  0.26810600E+37 &  0.45404400E+13 &  0.42918160E+31\\
  0.29776600E+37 &  0.50436140E+13 &  0.48089980E+31\\
  0.32742200E+37 &  0.55468420E+13 &  0.53594420E+31\\
  0.35707300E+37 &  0.60501410E+13 &  0.59429390E+31\\
  0.38672000E+37 &  0.65534760E+13 &  0.65589600E+31\\
  0.41636500E+37 &  0.70568810E+13 &  0.72068480E+31\\
  0.44600500E+37 &  0.75603230E+13 &  0.78857870E+31\\
  0.48552100E+37 &  0.82316740E+13 &  0.88379610E+31\\
  0.51515000E+37 &  0.87352400E+13 &  0.95862580E+31\\
  0.54477600E+37 &  0.92388600E+13 &  0.10362880E+32\\
  0.57439900E+37 &  0.97425150E+13 &  0.11167080E+32\\
  0.60401800E+37 &  0.10246220E+14 &  0.11998160E+32\\
  0.65337700E+37 &  0.11085860E+14 &  0.13440860E+32\\
  0.69285800E+37 &  0.11757640E+14 &  0.14645200E+32\\
  0.73233400E+37 &  0.12429500E+14 &  0.15892440E+32\\
  0.77180500E+37 &  0.13101440E+14 &  0.17181270E+32\\
  0.82113400E+37 &  0.13941490E+14 &  0.18848650E+32\\
  0.86059200E+37 &  0.14613590E+14 &  0.20226200E+32\\
  0.89018100E+37 &  0.15117700E+14 &  0.21284120E+32\\
  0.93948300E+37 &  0.15958020E+14 &  0.23093300E+32\\
  0.98877600E+37 &  0.16798420E+14 &  0.24958070E+32\\ 
\botrule
\end{tabular}}
\end{table}
As mentioned in Sec.~3.2.3, 
at subnuclear densities all the EoS considered here are joined with the crustal equations of state from Harrison and Wheeler\cite{HW} and Negele and Vautherin.\cite{NV}  
The composition of the crust is crystalline, with light\cite{HW} or heavy\cite{NV} metals and electron gas.
The DBHF equations of state for neutron matter and for $\beta$-equilibrated matter are given in Table~3 and
Table~4, respectively. Some points of the EoS used for the crust are displayed in Table~5. 

\begin{figure}[!t] 
\centering          
\vspace*{-1.2cm}
\includegraphics[totalheight=3.5in]{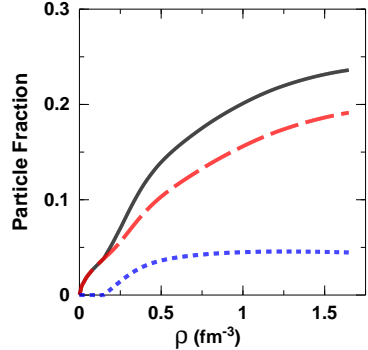} 
\vspace*{-2.2cm}
\caption{(color online) Proton (solid black), electron (dashed red), and muon (dotted blue) fractions
in $\beta$-stable matter as a function of total baryon density as predicted by the DBHF model. 
} 
\label{pfrac}
\end{figure}

The proton fraction in $\beta$-stable matter is calculated by imposing energy conservation and charge 
neutrality. The resulting algebraic equations can be found in standard literature.\cite{Glen}  
The contribution to the energy density from the electrons is written as      
\begin{equation} 
e_e= \frac{\hbar c}{4 \pi ^2}(3 \pi ^2 \rho _e)^{4/3} \, , 
\label{el}  
\end{equation}
whereas for muons we write     
\begin{equation} 
e_{\mu}= \rho _{\mu} m_{\mu}c^2 + (\hbar c)^2\frac{(3 \pi ^2 \rho_{\mu})^{5/3}}{10 \pi ^2 m_{\mu}c^2} \, .                   
\label{emu}  
\end{equation}
These contributions are added to the baryonic part to give the total energy density. 
The derivative of the total energy/particle with respect to the fraction of a particular particle species 
is the chemical potential of that species. The conditions 
\begin{equation} 
\mu _p + \mu _e = \mu _n \, \, ; \, \,                                                                                     
\mu _{\mu} = \mu _e  \, \,  ; \, \,                                                                                      
\rho _p = \rho _{\mu} + \rho _e \; , 
\label{beta}  
\end{equation}
allow to solve for the densities (or fraction) of protons, electrons, and muons. 
Near the saturation density, when the muon fraction is close to zero, one can estimate the equilibrium 
proton fraction, $x_p$, to be\cite{Latt07} 
\begin{equation} 
x_p \approx \Big (\frac{4 e_{sym}(\rho_0)}{\hbar c} \Big )^3/(3 \pi ^2 \rho _0). 
\label{xp}  
\end{equation}
The fractions of protons, electrons, and muons as predicted by DBHF are shown in Fig.~\ref{pfrac}. 
The critical density for the proton fraction to exceed approximately $1/9$ and, thus, allow cooling through the  direct Urca  processes,                          
\begin{equation} 
n \rightarrow p + e + \bar{\nu}_e \, \, \, \, \, and \, \, \, \, \, 
 p +e \rightarrow n + {\nu}_e \, ,                           
\label{Urca}  
\end{equation}
is about 
$0.36-0.39$ fm$^{-3}$. 

\begin{figure}[!t] 
\centering          
\includegraphics[totalheight=2.5in]{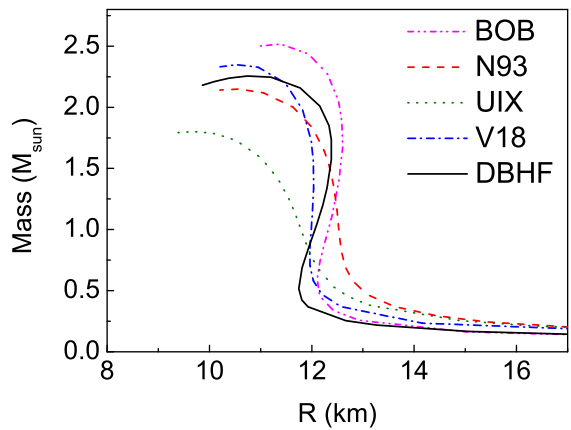} 
\vspace*{0.2cm}
\caption{(color online) Static neutron star mass-radius relation for the models
considered in the text. 
} 
\label{MR}
\end{figure}

In Fig.~\ref{MR}, we show the mass-radius relation for a sequence of static neutron stars as predicted
by the various models considered previously, see Fig.~\ref{esmicro}.                                    
\footnote{All neutron star properties are   
calculated from public software downloaded from the website {\it http://www.gravity.phys.uwm.edu/rns}.} 
All models besides DBHF share the same many-body
approach (BHF+TBF) but differ in the two-body potential and TBF employed. The resulting differences can be 
much larger than those originating from the use of different many-body approaches. This can be seen by comparing
the DBHF and BOB curves, both employing the Bonn B interaction (although in the latter case the non-reltivistic, r-space version of the potential is adopted). Overall, the maximum masses range from 1.8$M_{\odot}$
(UIX) to 
2.5$M_{\odot}$ (BOB). Radii are less sensitive to the EoS and range between 10 and 12 km for all models under
consideration, DBHF or BHF+TBF. 
Concerning consistency with present constraints, an initial observation of a neutron-star-white dwarf binary system
suggested a neutron star mass (PSR J0751+1807) of 2.1$\pm$0.2$M_{\odot}$.\cite{Nice} Such observation 
would imply a considerable constraint on the high-density behavior of the EoS. On the other hand,    
a dramatically reduced value             
of 1.3$\pm$0.2$M_{\odot}$                                          
was reported\cite{Piek08} later, which does not invalidate any of the theoretical models under
consideration. 

\begin{figure}[!t] 
\centering          
\includegraphics[totalheight=1.8in]{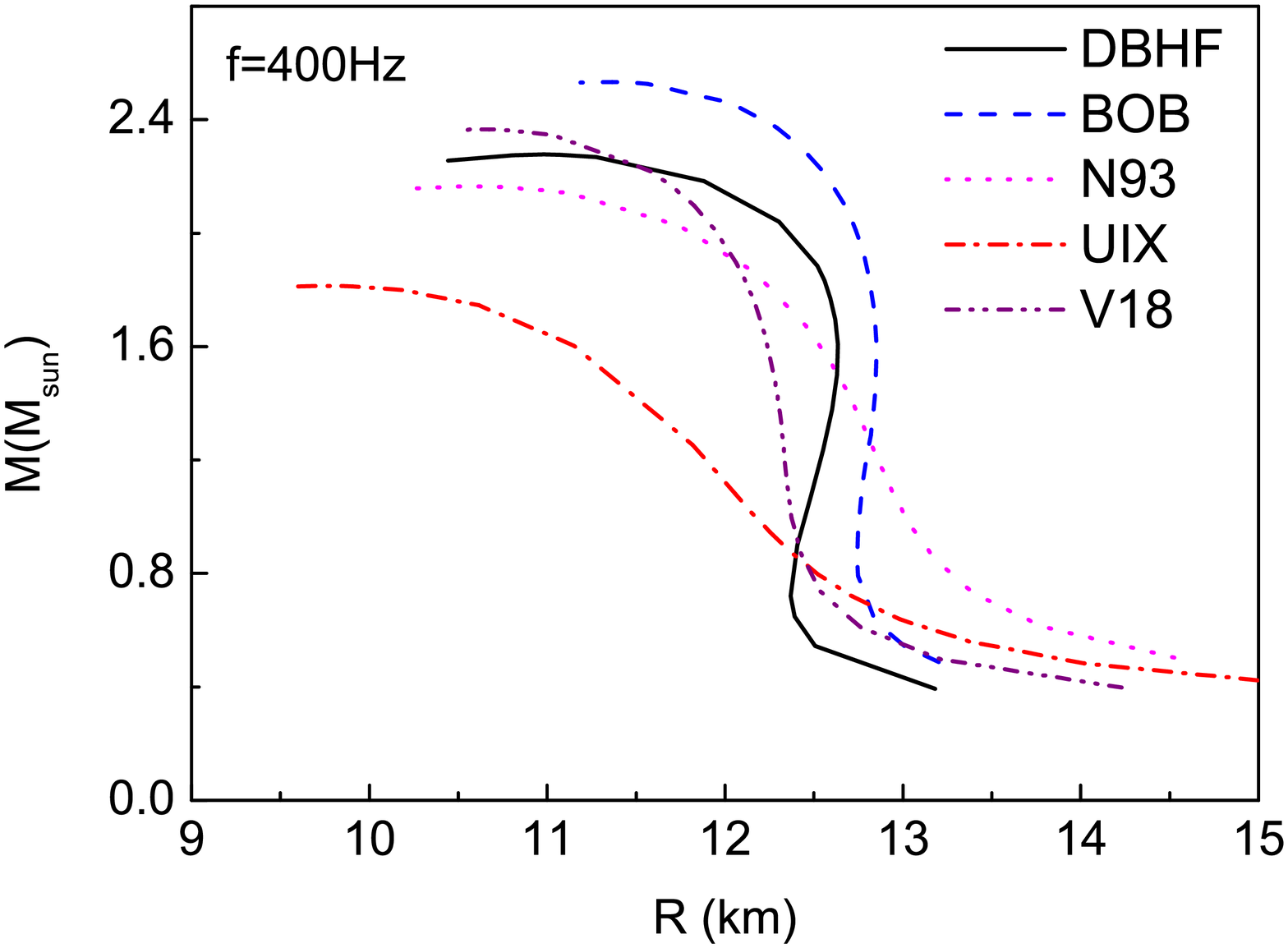}  
\includegraphics[totalheight=1.8in]{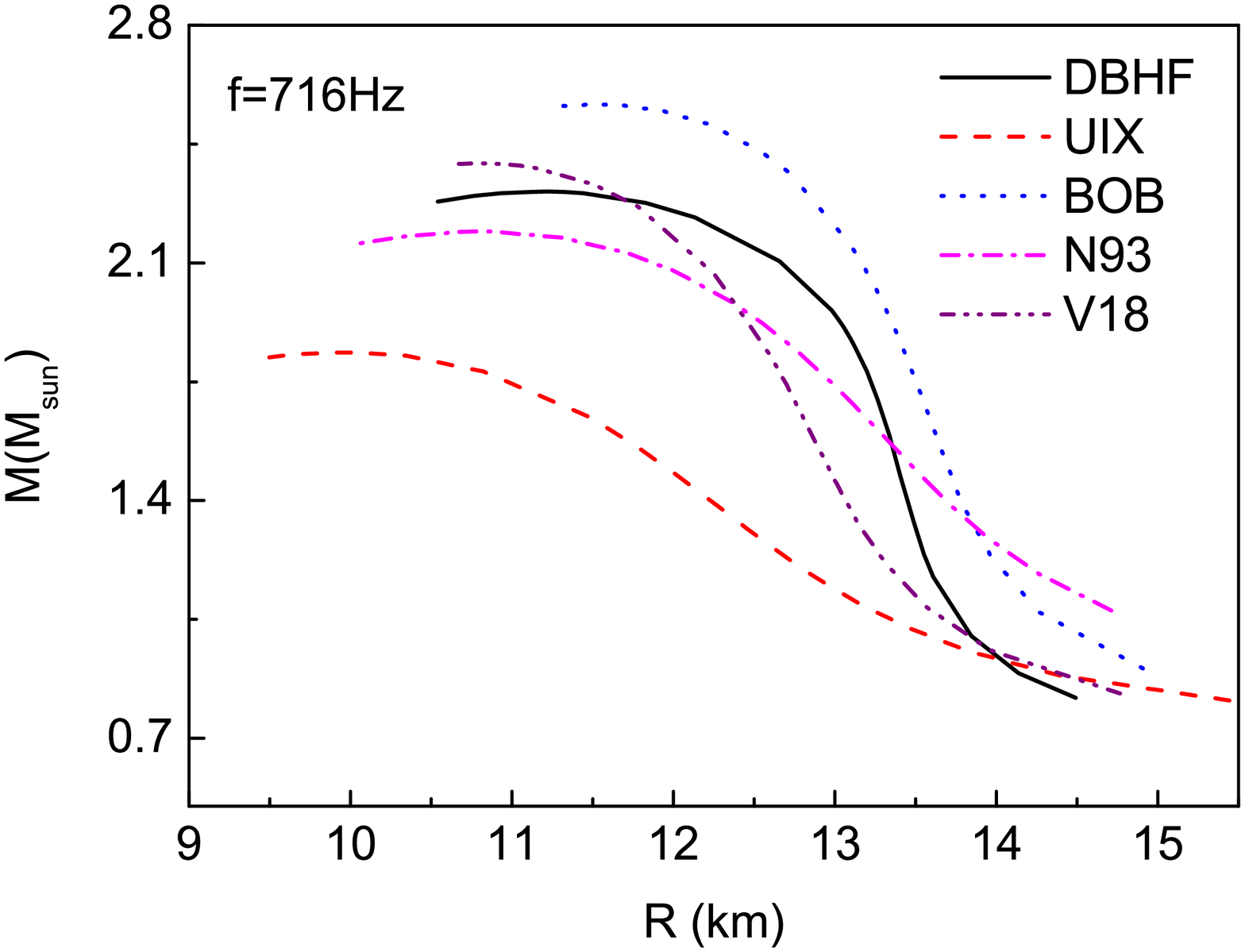}  
\includegraphics[totalheight=1.8in]{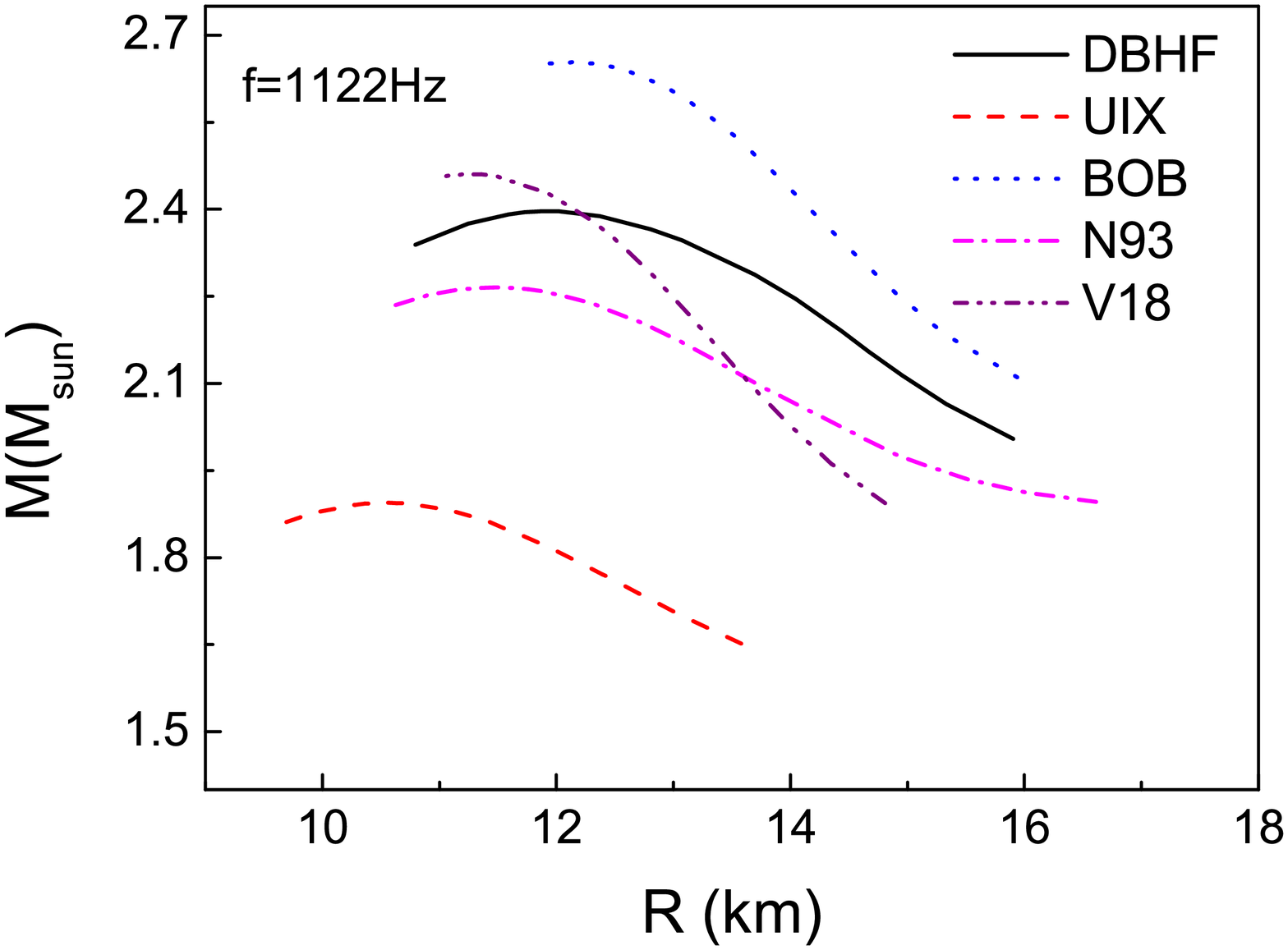}    
\vspace*{0.2cm}
\caption{(color online)  Mass-radius relation for the models
considered in the text and for different rotational frequencies.                  
} 
\label{MRrot}
\end{figure}

The model dependence is shown in Fig.~\ref{MRrot}  for the case of rapidly rotating stars. 
The 716 Hz frequency corresponds to the most rapidly rotating pulsar, PSR J1748-2446,\cite{Hessels}  
although recently an X-ray burst oscillation at a frequency of 1122 Hz has been reported\cite{Kaaret}
which may be due to the spin rate of a neutron star.
As expected, the maximum mass and the            
(equatorial) radius become larger with increasing rotational frequency.

\begin{figure}[!t] 
\centering          
\includegraphics[totalheight=1.8in]{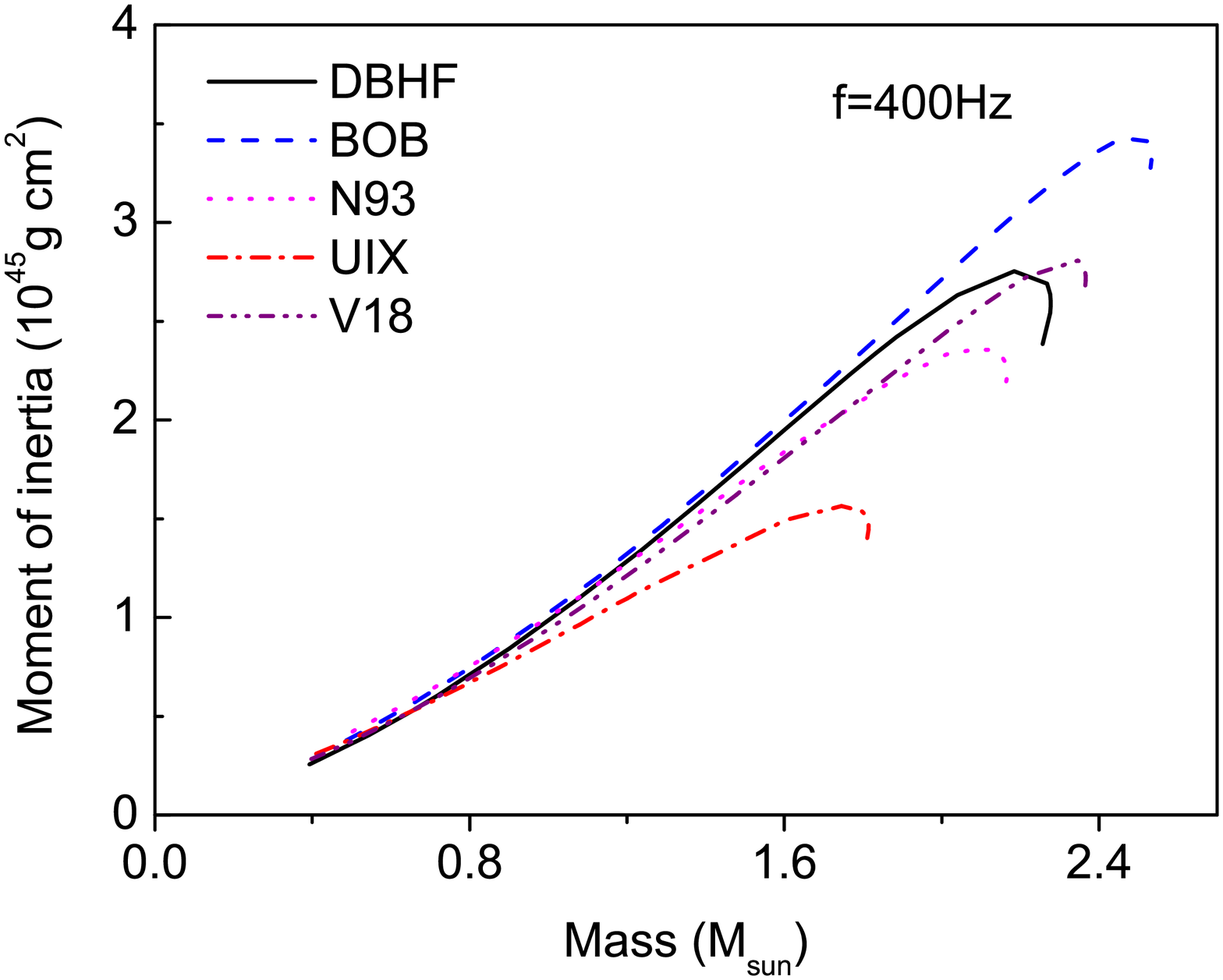}  
\includegraphics[totalheight=1.8in]{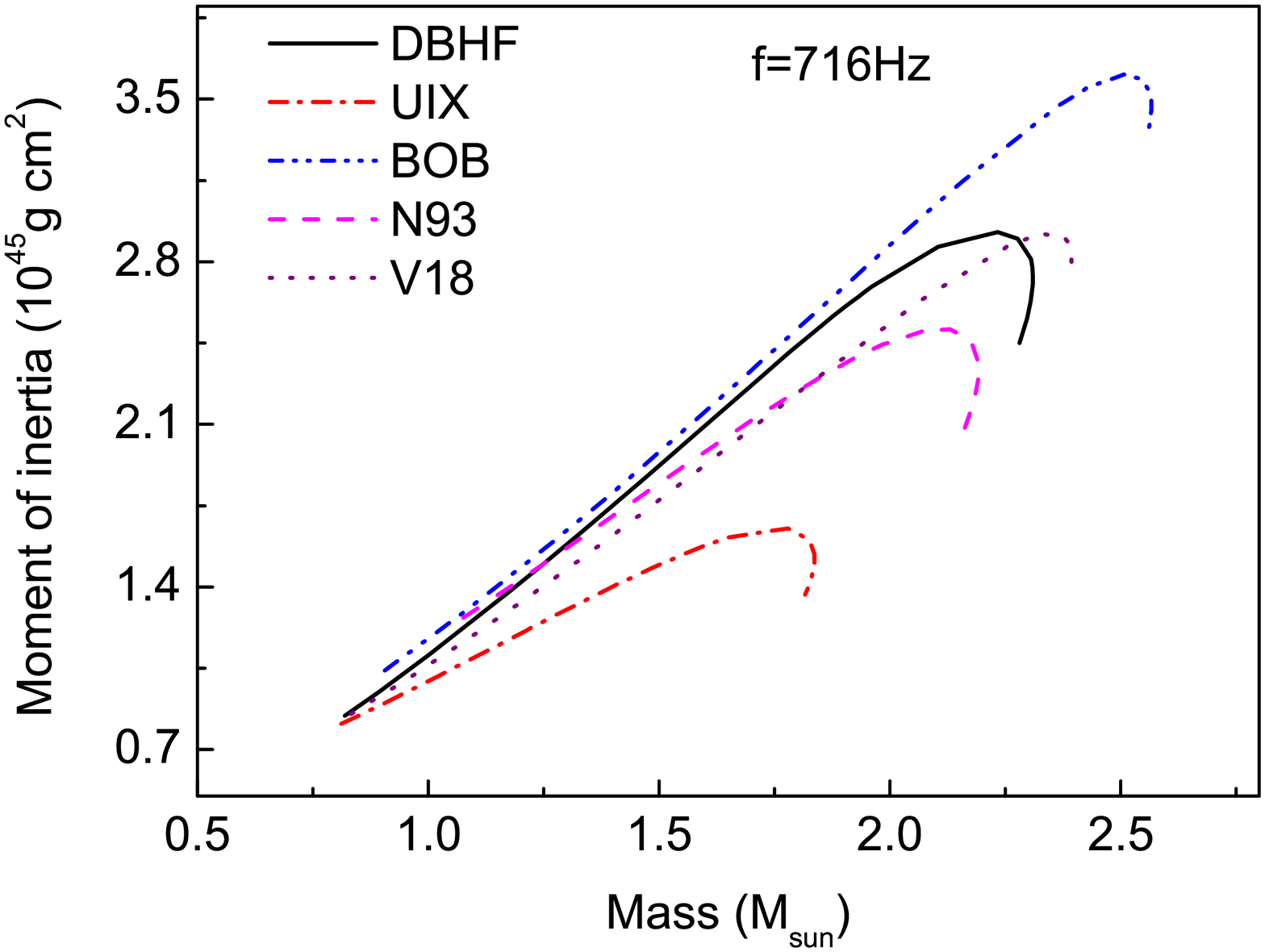}  
\includegraphics[totalheight=1.8in]{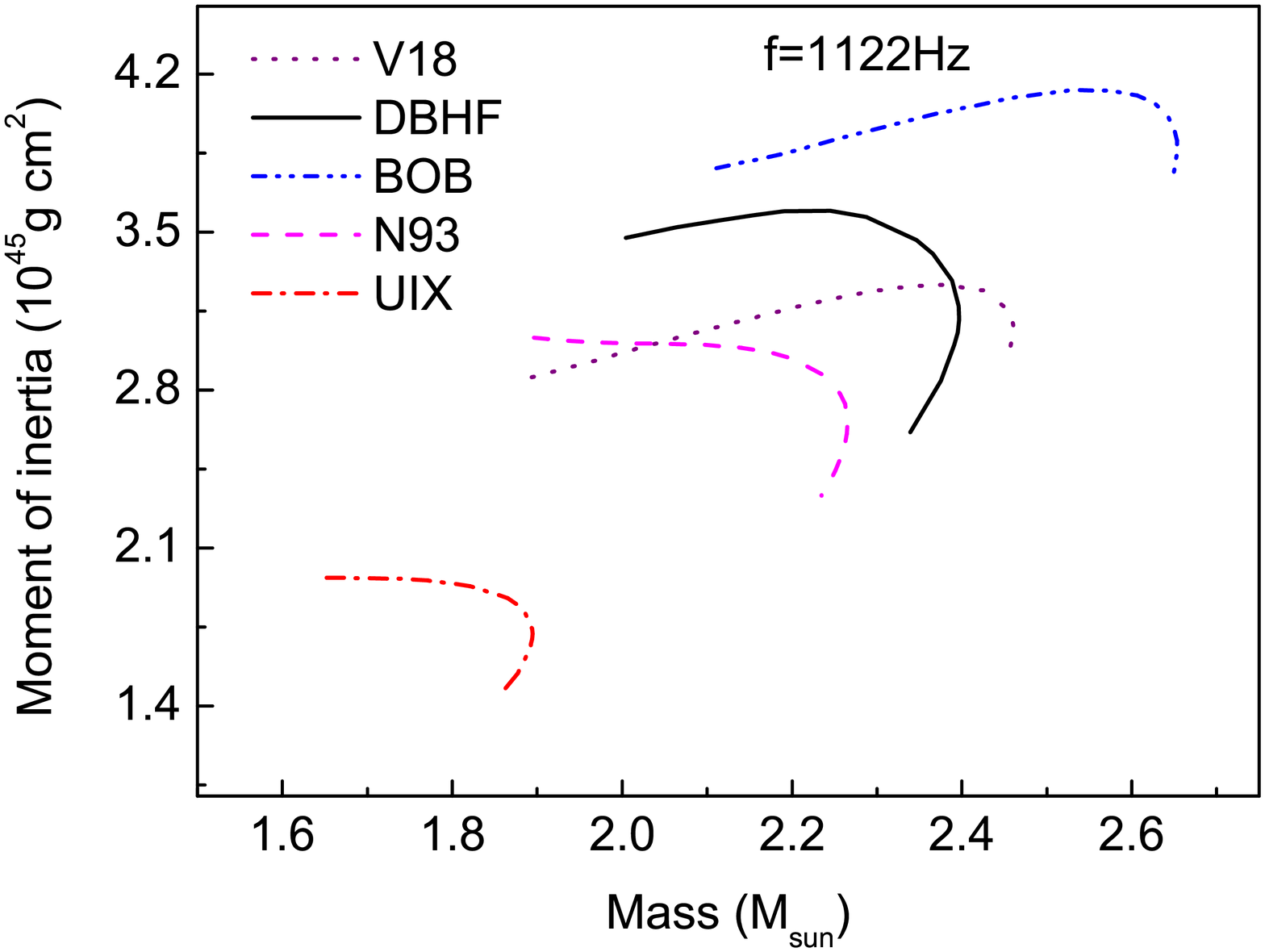}   
\vspace*{0.2cm}
\caption{(color online)  Moment of inertia for  the models
considered in the text and for different rotational frequencies.                  
} 
\label{IM}
\end{figure}

\begin{figure}[!t] 
\centering          
\includegraphics[totalheight=2.5in]{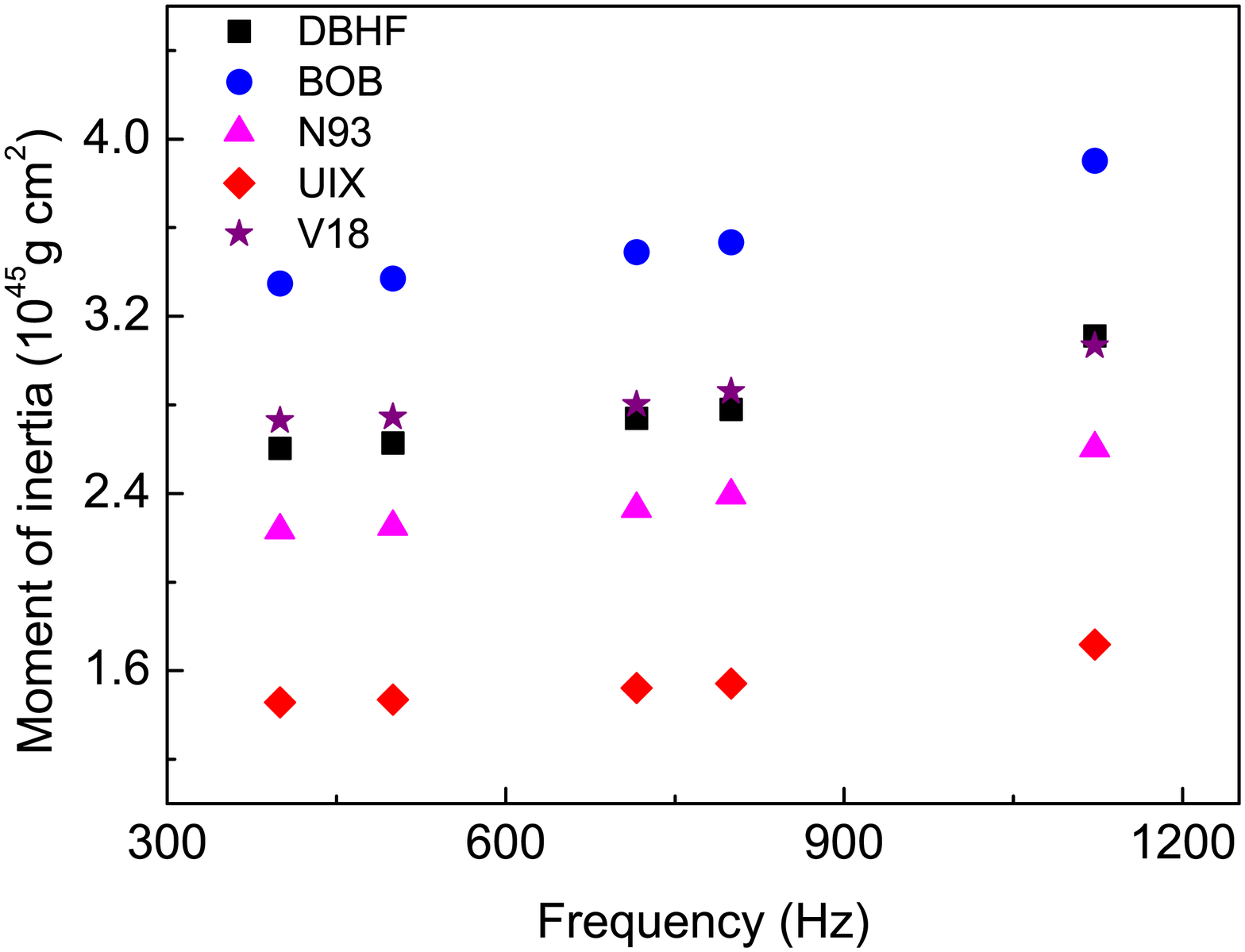}  
\vspace*{-0.5cm}
\caption{(color online)  Moment of inertia corresponding to the maximum mass for  the models
considered in the text as a function of the rotational frequency.                    
} 
\label{IF}
\end{figure}

Another bulk property of neutron stars is the moment of inertia, $I$. 
In Fig.~\ref{IM}, we show the moment of inertia at different rotational speeds (again, for all models), whereas in 
Fig.~\ref{IF} we display the moment of inertia corresponding to the maximum mass at different
rotational frequencies.
These values are not in contradiction with observations of the 
Crab nebula luminosity. From that, a lower bound on the moment of inertia was inferred to be 
$I \geq  $4-8 $\times$ 10$^{44}$ g cm$^2$, see Ref.\cite{Weber} and references therein. 
The size of $I_{M_{max}}$ changes from model to model in line with the size of the maximum mass,
see Fig.~\ref{IF}. 

Lastly, we calculate the gravitational redshift predicted by each model. 
The redshift is given by     
\begin{equation}
z = \Big (1 - \frac{2M}{R} \Big ) ^{-1/2} -1 \; . 
\label{red}
\end{equation}
This simple formula can be derived considering a photon emitted at the surface of a neutron star and moving
towards a detector located at large distance.\cite{Weber} The photon frequency at the emitter (receiver) is 
the inverse of the proper time between two wave crests in the frame of the emitter (receiver).   
Assuming a static gravitational field, and writing  $g_{00}$ as the metric tensor component 
at the surface of a nonrotating star yield the equation above. 
Naturally the rotation of the star modifies the metric, and in that case different considerations need to
be applied which result in a frequency dependence of the redshift. We will not consider the general case
here. 

\begin{figure}[!t] 
\centering          
\includegraphics[totalheight=2.5in]{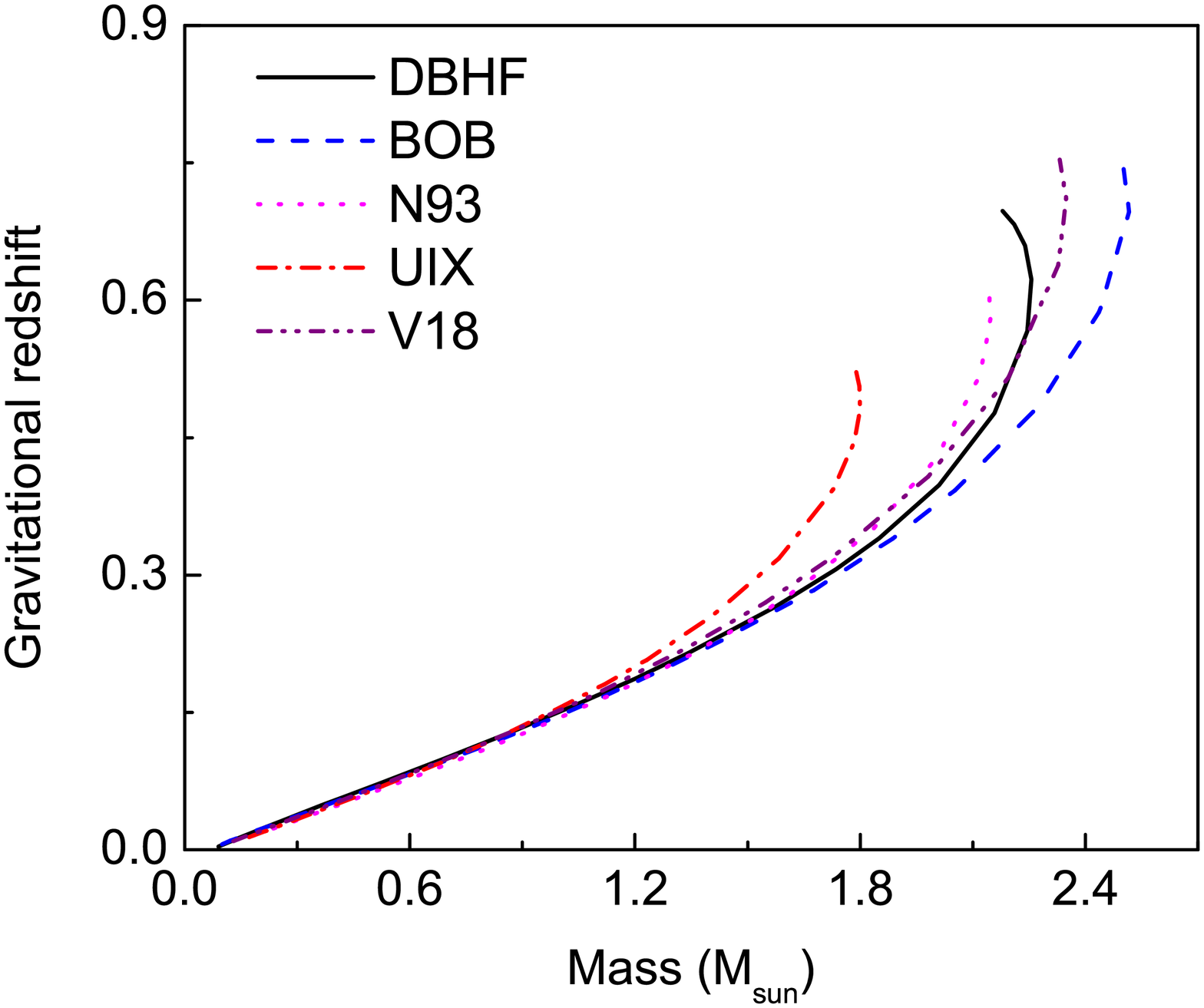}                
\vspace*{-0.5cm}
\caption{(color online)  Gravitational redshift for all models. 
For each model, the corresponding sequence of static stars is considered. 
} 
\label{Z}
\end{figure}

From Fig.~\ref{Z}, it appears that the gravitational redshift is not very EoS-dependent (compare, for instance, 
the values at the maximum mass for each model), an indication that the gravitational profile at 
the surface of the star is similar in all models.               
Thus measurements of $z$ may not be the best way to discriminate among different EoS. 

Clearly, at the densities           
probed by the interior of neutron stars the model dependence is large, but presently available constraints are 
still insufficient to discriminate among these EoS. 
At very high density ($\rho$ between five and ten times normal density), the most repulsive {\it 
symmetric matter} energies are produced with BOB, V18, DBHF, N93 and UIX, in that order.
This explains the maximum mass predictions, which depend mostly on the absolute repulsion
present in the symmetric matter EoS. 
In pure neutron matter, N93 follows right after BOB (again, from largest to smallest repulsion). This indicates
a somewhat different balance of attraction/repulsion when only T=1 contributions are 
included. 
The symmetry energy, see Fig.~\ref{esmicro}, which depends entirely on the repulsion of neutron matter {\it relative} 
to symmetric matter and 
whose density dependence controls observables such as the neutron skin, 
is largest in N93, followed by BOB, V18, UIX, and DBHF.

The model dependence we observe comes from two sources, the two-body potential and the 
many-body approach, specifically the presence of explicit TBF or Dirac effects. 
The dependence on the two-body potential is very large. Typically, the main source of model dependence 
among NN potentials is found in the strength of the tensor force.                                            
Of course, differences at the two-body level impact the TBF as well, whether they are microscopic
or phenomenological (as in the case of UIX). 

On the other hand, 
when comparing DBHF and BOB, we are looking at differences stemming from the many-body scheme, as          
the two models share the same NN potential. 
In the BOB model, repulsion grows at a much faster rate than in DBHF, and more strongly so in neutron 
matter. (Hence, the much larger symmetry energy with BOB). As an example, at about 6 times
normal density the DBHF energy of symmetric matter is 67\% of the BOB energy and only 51\% at          
ten times normal densities. In pure neutron matter, those ratios become 49\% and 30\%, respectively. 
Thus, the inclusion of the microscopic TBF in the BHF model introduces considerable more 
repulsion than the Dirac effects                                                                      
through highly non-linear                  
terms.\cite{Catania3}                                       
Both attractive and repulsive TBF are required for a realistic description of the 
saturation point. The density dependence of the repulsive terms is obviously stronger and thus 
dominates at high density. 
Furthermore, it appears that this is especially true in neutron matter. 

In Ref.\cite{Catania3} it is shown that the largest contribution to the net TBF originates from Z-diagrams such
as shown in Fig.~2, a fact which confirms the validity of the DBHF approximation. The actual amount of repulsion, 
however, seems to depend sensitively upon the TBF parametrization. One of the advantages of the DBHF method
is the ability to avoid possible inconsistencies between the parameters of the two- and the three-body systems. 

At this time, available constraints cannot pin down the high-density behavior of the EoS. 
Nevertheless, we argue again that microscopic models allow for a deeper insight into the            
origin of the observed physical effects, 
and should be pursued along with improved constraints.

\section{Non-Nucleonic Degrees of Freedom}

\subsection{Introduction}
                                                                     
At densities close to normal nuclear density, protons and neutrons are the only baryonic degrees
of freedom. As density increases, other baryons begin to appear, such as strange baryons 
or isospin 3/2 nucleon resonances. 
 Hyperonic states can be classified according to the irreducible representation of the 
$SU(3)$ group. The octet of baryons that can appear in neutron matter includes nucleons,
$\Lambda$, $\Sigma ^{0,\pm}$, and $\Xi ^{0,-}$. 

Neglecting the nucleon-hyperon interaction, the threshold for stable hyperons to exist in matter is       
determined by comparing the hyperon mass with the neutron Fermi energy, which is the largest available energy scale 
in neutron-rich matter.                                                                                         
We consider cold neutron stars, after neutrinos have escaped. Strange baryons appear at about 2-3 times
normal density,\cite{nycatania2} an estimate which is essentially model independent, through the
processes $n + n \rightarrow p + \Sigma ^-$ and 
 $n + n \rightarrow n + \Lambda$.      
The equilibrium conditions for these reactions are
\begin{equation} 
2 \mu _n = \mu _p + \mu _{\Sigma^-} \, \, ; 
\, \, \mu _n = \mu _{\Lambda}\, .                     
\label{nybeta1} 
\end{equation} 
Also, we have 
\begin{equation} 
 \mu _e = \mu _{\mu} \, \, ; 
\, \, \mu _n = \mu _p + \mu _e \, ,                     
\label{nybeta1b} 
\end{equation} 
the equations above being special cases of 
\begin{equation} 
 \mu  = b\mu _{n} - q \mu _e \, , 
\label{nybeta1c} 
\end{equation} 
where $b$ and $q$ are the baryon number and the charge (in units of the electron charge)
of the particular species with chemical potential $\mu$. 
Together with the charge neutrality condition and baryon number conservation,
\begin{equation} 
\rho _p = \rho _e + \rho _{\mu} + \rho _{\Sigma ^-} \, \, ; 
\, \, \rho  = \rho _n + \rho _p + \rho _{\Sigma ^-}              
+ \rho _{\Lambda}\, ,        
\label{nybeta2} 
\end{equation} 
the above system allows to determine the various particle fractions.

Clearly, 
the composition of matter at supra-nuclear densities determines the behavior of stellar matter.
It is also speculated that a transition to a quark phase may take place at very high densities, 
the occurrence of which depends sensitively on the properties of the EoS in the hadronic (confined)
phase.                 
The presence of hyperons in the interior of neutron stars is reported to 
soften the equation of state, with the consequence that the predicted             
neutron star maximum masses become considerably smaller.\cite{Sch+06} With recent 
constraints allowing maximum masses larger than previously accepted limits (see
previous section), 
accurate microscopic calculations which include strangeness (in addition to other  
important effects, such as those originating from relativity), 
become especially timely. 

Thus, 
there are strong motivations for including strange baryons in nuclear matter.
Moreover, as far as terrestrial nuclear physics is concerned, studies 
of hyperon energies in nuclear matter naturally complement our knowledge
of hypernuclei (see, for instance, Refs.\cite{hyper1,hyper2,hyper3,hyper4}). 
For example, the EoS of hypermatter is 
useful in the development of generalized mass formulas depending on both density and hyperon fraction.\cite{hyper4,nycatania1}  
From the experimental side, additional data are very much needed, especially
on $\Lambda \Lambda$ hypernuclei, which would provide information on the 
$\Lambda- \Lambda$ interaction. Concerning single hypernuclei, analyses of              
data on $\Lambda$ binding energies constrain 
the depth of the single-$\Lambda$ potential to be 27-30 MeV.\cite{Dover}  
The status of $\Sigma$ hypernuclei and the $\Sigma$-nucleus potential is more controversial 
(see Ref.\cite{Saha04} and references therein).                 

\subsection{Hyperons and neutron star matter}

Interacting hypernuclear matter was initially studied within variational approaches.\cite{Pan,BJ}  
Relativistic mean field models were also extended to include hyperons.\cite {yn1,yn2,yn3,yn4} 
Microscopic models require 
inclusion of realistic nucleon-hyperon and hyperon-hyperon interactions,\cite{NY89,Nij89,Rij99,Stocks99} 
but the experimental information on these interactions is still scarce. The nucleon-hyperon
potentials are fitted to $\Lambda N$ or $\Sigma N$ scattering data. The information on the
hyperon-hyperon interaction is limited to the ground state of double-$\Lambda$ hypernuclei.\cite{Gib94} 
Additional constraints can be derived from $SU(3)$ symmetry arguments. 

The common denominator among microscopic 
calculations of the EoS with hyperons is that they are typically conducted within a   
non-relativistic framework together with r-space local nucleon-hyperon (NY) potentials.                       
Microscopic calculations of nuclear matter properties including hyperons have been reported earlier  
within the non-relativistic Brueckner-Hartree-Fock framework (BHF) (see, for instance, 
Refs.\cite{nycatania2,nycatania1}), using             
the Nijmegen\cite{Nij89}         
nucleon-hyperon (NY) meson-exchange potential.                                     
Extensive microscopic work on hyperonic nuclear matter, again within the non-relativistic BHF framework, has also been published by 
the Barcelona group (see, for instance, Refs.\cite{Vid+00,Vid+00b,Vid+04,Rios+05}).

The issue of interacting hyperons and their impact on the $\beta$-stable EoS and neutron star structure
has been confronted, for instance, in Ref.\cite{nycatania3} using the Paris\cite{Paris} and the Argonne V18\cite{V18} NN potentials together with the Nijmegen soft-core
(NSC89) NY potential.\cite{Nij89}
A remarkable conclusion from that work is that, in the presence of hyperons, the inclusion of nucleonic
TBF does not alter the EoS appreciably. With nucleons only,
TBF bring in considerable repulsion at high density which result in a much stiffer EoS. On the other hand, 
when hyperons are present the nucleonic TBF increase the strange baryon population due to the 
increased nucleon chemical potential. In turn, this decreases the nucleon population with the final net effect 
on the EoS found to be very small. In other words, it would be necessary for the NY interaction to become         
very repulsive at high density to compensate for the loss of nucleons and gain a substantial increase of the star maximum   
mass.\cite{nycatania3}
 Results for the neutron star maximum gravitational mass, radius, and central 
density are shown in Table~6. 
The use of different NN potentials seem to produce only small variations.           

\begin{table}[pt]
\centering \caption                                                             
{Neutron star limiting values for different EoS with (Y) and without (no Y) hyperons as 
 from Ref.$^{132}$.                           
}       
\vspace{5mm} 
\begin{tabular}{ccccccc}     
\hline 
\multicolumn{1}{c}{EoS} & 
\multicolumn{2}{c}{$M_{max}/M_{\odot}$} &\multicolumn{2}{c}{$R (km)$} & \multicolumn{2}{c}
                   {$\rho _c (fm ^{-3})$} \\               
      & no Y & Y& no Y&  Y&  no Y& Y \\ \colrule 
V18 & 1.64 & 1.26 & 9.10 & 8.70 & 1.53 & 1.86 \\                
Paris & 1.67 & 1.31 & 8.90 & 8.62 & 1.59 & 1.84 \\                
V18+TBF & 2.00 & 1.22 & 10.54 & 10.46 & 1.11 & 1.25 \\                
Paris+TBF & 2.06 & 1.26 & 10.50 & 10.46 & 1.10 & 1.25 \\                
\hline 
\end{tabular} 
\end{table}

More recently,                                                                                     
 neutron star structure results                  
have been revisited by the Catania group 
with the microscopic models we considered earlier in this article (BHF+TBF models from Ref.\cite{Catania4})           
along with the NSC89 nucleon-hyperon interaction.\cite{Nij89}  
There, the nucleonic energy densities obtained with the different NN interactions are used together with 
the same NY interaction to calculate stellar matter. It is found that the inclusion of hyperons reduces dramatically the maximum mass. 
It also reduces substantially the maximum mass range covered by the models, which goes from 1.8 to 
2.5~$M_{\odot}$ with nucleonic energy densities only, see Fig.~\ref{MR}, to the much narrower intervals of 1.3-1.4~$M_{\odot}$. 
Furthermore, the stiffer the original EoS, the larger the softening effect from the presence of strangeness. 
Thus, in this approach, one may conclude that hybrid stars (containing hadronic and quark matter) are
necessary to reproduce the larger mass values consistent with recent observations.

Overall, there seems to be a consensus among microscopic calculations of the BHF type (with or 
without TBF) that the inclusion of hyperons yields a remarkably low value of a neutron star maximum
mass. On the other hand, predictions from different versions of relativistic mean field models are quite            
different from one another.\cite{RMF1,RMF2}                                                     

The alternative approach to the EoS with strange baryons presented in the next section 
will suggest that       
present NY interactions are not yet sufficiently constrained to allow for definite conclusions 
concerning the occurrence of a hybrid phase.

\subsection{A first DBHF calculation of the EoS with nucleons and $\Lambda$-hyperons}

It is one purpose of this section to bring in the new aspect of Dirac effects 
on the $\Lambda$ hyperon as well as the nucleon. By ``Dirac effects" we mean that the single-baryon            
 wavefunction is calculated self-consistently with the appropriate effective interaction.
The origin and nature of these effects on the nucleonic equation of state was discussed previously,                
see Sec.~3.2.1. 

A previous calculation from the Idaho group\cite{Sam08} of the binding energy of a 
$\Lambda$ impurity in nuclear matter showed that Dirac effects on the $\Lambda$ hyperon               
yield a moderate reduction of the binding energy. In that calculation, we used the most recent
nucleon-hyperon (NY) potential reported in Ref.\cite{NY05} (thereafter referred to as NY05), and observed
that $G$-matrix predictions obtained with NY05 are                                  
significantly differerent from calculations using the previous 
(energy independent) version 
of the J{\"u}lich NY potential\cite{NY94}.                                                              
Therefore, in this work we will use both potentials, for comparison.                       
As usual, the Bonn B potential \cite{Mac89} is used throughout for the nucleon-nucleon (NN) part.

Previous calculations of the EoS with hyperons have typically been conducted within a   
non-relativistic framework together with r-space local NY potentials. The DBHF calculation we describe here uses     
non-local relativistic momentum-space (NN and NY) potentials and a relativistic many-body 
method, 
and is therefore fundamentally different.                          
Next we give a brief review of the formalism. 
                 
The single-nucleon and single-$\Lambda$ potentials are obtained as 
\begin{equation}
U_N({\vec k_N}) = U_{N\Lambda} ({\vec k_N}) + U_{NN}({\vec k_N}),                    
\label{UN}
\end{equation}
and 
\begin{equation}
U_{\Lambda}({\vec k}_{\Lambda}) = U_{\Lambda N} ({\vec k}_{\Lambda}),                                  
\label{ULAM}
\end{equation}
i.e., the $\Lambda \Lambda$ interaction is neglected.  
In the equations above, the various terms, $U_{B_1B_2}$, are the contributions to the potential 
of baryon $B_1$ from its interation with all baryons of type $B_2$. They are given by
\begin{equation}
 U_{N\Lambda} ({\vec k_N}) = \sum_{T,L,S,J} \frac{(2T+1)(2J+1)}{(2t_N+1)(2s_N+1)}                               
  \int _0 ^{k_F^{\Lambda}} G^{T,L,S,J}_{N \Lambda}(k({\vec k_N},{\vec k_{\Lambda}}),  
  P({\vec k_N},{\vec k_{\Lambda}})) d^3 k_{\Lambda}, 
\label{UNL}
\end{equation}
\begin{equation}
 U_{N N} ({\vec k_N}) =                                                                  
 \sum_{T,L,S,J} \frac{(2T+1)(2J+1)}{(2t_N+1)(2s_N+1)}                               
 \int _0 ^{k_F^{N}} G^{T,L,S,J}_{N N}(k({\vec k_N},{\vec k_{N}'}),  
 P({\vec k_N},{\vec k_{N}'})) d^3 k_{N}', 
\label{UNN}
\end{equation}
and 
\begin{equation}
 U_{\Lambda N} ({\vec k}_{\Lambda}) =                                                                            
 \sum_{T,L,S,J} \frac{(2T+1)(2J+1)}{(2t_{\Lambda}+1)(2s_{\Lambda}+1)}                               
 \int _0 ^{k_F^{N}} G^{T,L,S,J}_{\Lambda N}(k({\vec k_N},{\vec k_{\Lambda}}),  
  P({\vec k_N},{\vec k_{\Lambda}})) d^3 k_{N}, 
\label{ULN}
\end{equation}
where the channel isospin $T$ can be 0 or 1 for the NN case and is equal to 1/2 for the 
$N\Lambda$ case, and $s_i,t_i$ ($i=N,\Lambda$) are the spin and isospin of the nucleon or
$\Lambda$. 

Notice that      
\begin{equation}
\frac{U_{N\Lambda}}{U_{\Lambda N}} \approx \frac{\rho _{\Lambda}}{\rho _N}, 
\label{ny10}
\end{equation}
an approximation often used in mean-field approaches. 

The average potential energies of nucleons and $\Lambda$'s are determined from 
\begin{equation}
<U_N> =\frac{1}{\rho_N}\frac{1}{(2 \pi)^3}4\frac{1}{2}\int_0^{k_F^N} U_{N} ({\vec k_N}) 
dk^3_N,          
\label{ny11}
\end{equation}
and 
\begin{equation}
<U_{\Lambda}> =\frac{1}{\rho_{\Lambda}}\frac{1}{(2 \pi)^3}2\frac{1}{2}\int_0^{k_F^{\Lambda}} U_{\Lambda } ({\vec k}_{\Lambda})
dk^3_{\Lambda},          
\label{ny12}
\end{equation}
where the factors of 4 and 2 in Eqs.~(\ref{ny11}) and Eq.~(\ref{ny12}), respectively, account for protons and neutrons in both spin states
or $\Lambda$'s in both spin states.

Finally the average potential energy per baryon is obtained as
\begin{equation}
<U> = \frac{\rho _N <U_N> + \rho _{\Lambda}<U_{\Lambda}>}{\rho _{tot}},
\label{ny13}
\end{equation}
from which, together 
with a similar expression for the kinetic energy, one obtains the average energy per 
baryon. 

The N$\Lambda$ $G$-matrix is obtained from the Bethe-Goldstone equation
\begin{eqnarray} 
<N\Lambda|G_{N \Lambda}(E_0)|N\Lambda> & = & <N\Lambda|V|N\Lambda>     \\ 
                                       &   & \mbox{} +\sum_{Y=\Lambda,\Sigma} <N\Lambda|V|NY> \frac{Q}{E_0 - E}
<NY|G_{N \Lambda}(E_0)|N\Lambda> , \nonumber         
\label{ny14}
\end{eqnarray}
where $E_0$ and $E$ are the starting energy and the energy of the intermediate NY state,  
respectively, and $V$ is an energy-independent NY potential.

For two particles with masses $M_N$ and $M_{\Lambda}$ and Fermi momenta
$k_F^N$ and $k_F^{\Lambda}$, Pauli blocking requires               
\begin{equation}
Q({\vec k},{\vec P}) = 
                \left\{
\begin{array}{l l}
 1          & \quad \mbox{$|\beta {\vec P}+{\vec k}|>k_F^{\Lambda}$  and 
	   $|\alpha {\vec P}-{\vec k}|>k_F^{N}$ }    \\            
 0          & \quad \mbox{otherwise.}          
\end{array}
\right.
\label{ny15}
\end{equation} 
The above condition implies the restriction 
\begin{equation}
\frac{(\frac{M_N}{M}P)^2 +k^2-(k_F^N)^2}{2Pk\frac{M_N}{M}} > cos \theta >
-\frac{(\frac{M_{\Lambda}}{M}P)^2 +k^2-(k_F^{\Lambda})^2}{2Pk\frac{M_{\Lambda}}{M}},                   
\label{ny16}
\end{equation} 
where $\theta$ is the angle between the total (${\vec P}$) and the relative
(${\vec k}$) momenta of the two particles, and $M=M_{\Lambda} + M_N$. Angle-averaging is then applied in the 
usual way. 

In the present calculation we consider a non-vanishing density of $\Lambda$'s but 
do not allow for the presence of real $\Sigma$'s in the medium 
(although both $\Lambda$ and $\Sigma$                        
are included in the coupled-channel calculation of the NY $G$-matrix, see Eq.~(\ref{ny14})).   
Essentially we are considering a scenario where a small fraction of nucleons is replaced
with $\Lambda$'s, as could be accomplished by an experiment aimed at producing multi-$\Lambda$ hypernuclei.
Multistrange systems, such as those produced in heavy-ion collisions, may of course contain other hyperons 
on the outset. 
For small $\Lambda$ densities, though, as those we consider here, the cascade ($\Xi$) and the $\Sigma$ hyperon
are expected to decay quickly through the strong processes 
 $N+\Xi \rightarrow \Lambda+\Lambda$ and 
 $N+\Sigma \rightarrow N+\Lambda$. Under these conditions, a mixture of nucleons and $\Lambda$'s 
can be considered ``metastable", in the sense of being equilibrated 
over a time scale which is long relative to strong processes. 
 (In fact, the strong reactions mentioned above would have to be 
Pauli blocked in order to produce a metastable multistrange system \cite{SG00}.) 

We neglect the $YY'$ interaction, as very little is known        
about it. Furthermore, non-local momentum-space $YY$ potentials, appropriate for our DBHF framework, 
are not available at this time.  
For these reasons, we keep the $\Lambda$ concentration           
relatively low. 

We have incorporated DBHF effects in the $\Lambda$ matter calculation, which amounts to involving the 
$\Lambda$ single-particle Dirac wave function in the self-consistent calculation through the 
$\Lambda$ effective mass, $M^*_{\Lambda}$.\cite{nysam}                 
However, a problem that needs to be addressed with the        
J{\"u}lich NY potential in conjunction with DBHF calculations                             
is the use of the pseudoscalar coupling for the interactions of 
pseudoscalar mesons (pions and kaons) with nucleons and hyperons. For the reasons 
described in Sec.~3.2.1 (that is, the close relationship between Dirac effects and ``Z-diagram"
contributions), 
this relativistic correction is known to become unreasonably large when applied to 
a vertex involving pseudoscalar coupling. On the other hand, the gradient (pseudovector) 
coupling (also supported by chiral symmetry arguments) largely suppresses antiparticle contributions. 
To resolve this problem, one can make use of the on-shell equivalence between the pseudoscalar and
the pseudovector coupling, which amounts to 
relating the coupling constants as follows: 
\begin{equation}
g_{ps} = f_{pv}\frac{M_i + M_j}{m_{ps}},                   
\label{ny19}
\end{equation} 
where $g_{ps}$ denotes the pseudoscalar coupling constant and $f_{pv}$ the pseudovector one;
$m_{ps}$, $M_{i}$, and $M_{j}$ are the masses of the                         
pseudoscalar meson and the two baryons involved
in the vertex, respectively. 
This procedure can be made plausible by                                                               
writing down the appropriate 
one-boson-exchange amplitudes and observing that, redefining the coupling constants as above, we have 
(see Ref.\cite{Mac89} for the two-nucleon case)                  
\begin{equation}
V_{pv} = V_{ps} + .....             
\label{ny20}
\end{equation} 
where the ellipsis stands for off-shell contributions. 
 Thus, the pseudoscalar coupling can be interpreted as pseudovector coupling where the 
off-shell terms are ignored. This is what we apply in our DBHF calculations.                  
More concretely, from a given pseudoscalar potential, first we extract the corresponding $f_{pv}$ from 
Eq.~(\ref{ny19}). In the medium, $V_{ps}^*$ will contain the coupling constant 
$g=\frac{M_i^* + M_j^*}{m_{ps}}f_{pv}$. When the baryon masses are reduced through the effective mass
prescription, the reduction of the masses appearing at the denominator of the momentum-dependent part of the potential will
be ``balanced" be the equally reduced masses at the numerator of the coupling constant. This 
prescription prevents the OBE diagram from growing to unreasonably large values when effective masses are employed. 

Next, we will be showing results for both the NY05 potential \cite{NY05} and the previous version of the 
J{\"u}lich NY potential, NY94, specifically the model referred to as ${\tilde A}$ 
in Ref.\cite{NY94}.

\begin{figure}
\begin{center}
\vspace*{-2.0cm}
\hspace*{-2.0cm}
\scalebox{0.4}{\includegraphics{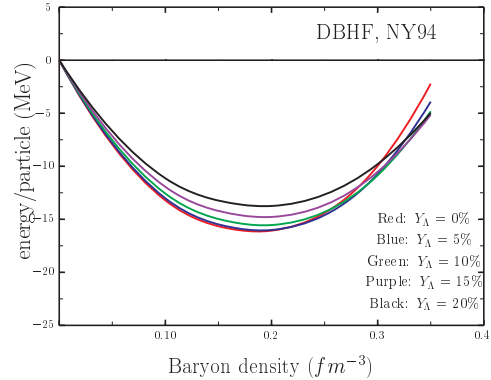}}
\vspace*{-2.6cm}
\caption{(Color online) Energy/particle as a function of density in symmetric nuclear matter for various
$\Lambda$ concentrations $Y_{\Lambda}$. Predictions obtained from DBHF calculations with the NY94 potential.
} 
\label{NY1}
\end{center}
\end{figure}

In Fig.~\ref{NY1} we show the energy per particle as a function of density for different $\Lambda$
concentrations as obtained from DBHF calculations along with the NY94 potential.                 
As more nucleons are replaced with 
$\Lambda$'s, generally less binding energy per particle is generated. This is due to the weaker nature of the
$N\Lambda$ interaction relative to the NN one. 
Furthermore, the $\Lambda$ Fermi momentum grows rather quickly with $\rho _{\Lambda}$, since only
two $\Lambda$'s can occupy each state, rather than four, which implies a fast rise of the 
hyperon kinetic energy. (Although, for small hyperon densities,                  
there is at first some reduction of the kinetic energy, due to the fact that          
$\Lambda$'s have larger mass and can occupy lower energy levels.)     
We notice that the saturation density remains essentially unchanged 
with increasing hyperon concentrations. 
As density grows, though, larger $\Lambda$ concentrations start to yield increased attraction.
One must keep in mind that, especially at the higher densities, the NN interaction become less and less
attractive due to medium effects, in particular repulsive Dirac effects. Thus, removing nucleons
from the system can actually amount to increased binding.

\begin{figure}
\begin{center}
\vspace*{-2.0cm}
\hspace*{-2.0cm}
\scalebox{0.4}{\includegraphics{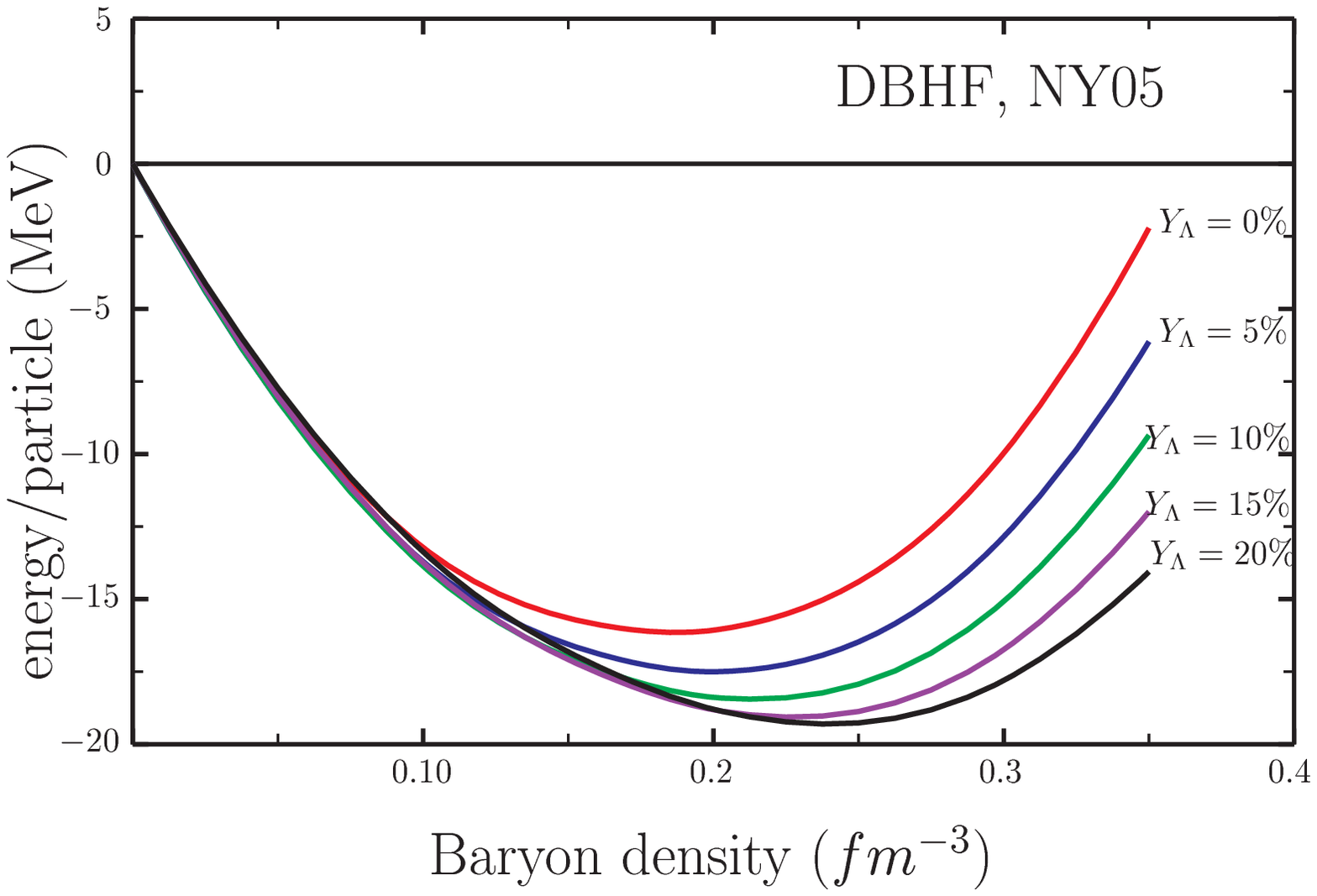}}
\vspace*{-3.0cm}
\caption{(Color online) Energy/particle as a function of density in symmetric nuclear matter for various
$\Lambda$ concentrations $Y_{\Lambda}$. Predictions obtained from DBHF calculations with the NY05 potential. 
} 
\label{NY2}
\end{center}
\end{figure}

Moving now to Fig.~\ref{NY2}, where the NY05 potential is adopted,                           
we see a very different scenario.                                    
We recall that the NY05 model is considerably more attractive, yielding
about 50 MeV for the $\Lambda$ binding energy \cite{Sam08,Pol} whereas a value close to 30 MeV was found with NY94         
\cite{NY94}. 
Naturally, we expect these differences to reflect onto the respective EoS predictions.
Here, increasing the hyperon population yields more binding, 
a trend opposite to the one seen in the previous figure.
Again, the final balance is the result of a combination of effects.
The fact that the NN component is repulsive at the higher densities, 
together with the more attractive nature of the NY05 potential, determines here a net increase in attraction 
with decreasing {\it nucleon} density. 
However, the additional binding becomes smaller and
smaller with increasing hyperon concentration, indicating that, at sufficiently large $\Lambda$
densities, the net balance may turn repulsive.             
Notice also that in Fig.~\ref{NY2} the minimum moves towards higher densities, signifying that
baryon pairs favor a smaller interparticle distance.

To summarize, one must keep in mind that the energy/particle is the result of a delicate balance          
of both the kinetic energies and the potential energies of the two baryon species, being weighed by
the respective densities. 
Thus, although the NN interaction is generally more attractive than the N$\Lambda$ one, 
the net effect of replacing  nucleons with $\Lambda$'s will depend 
sensitively on the nature of the NY interaction that's being put into the system, as well as the 
``stiffness" of the original, nucleonic, EoS. 

\begin{figure}
\begin{center}
\vspace*{-2.5cm}
\hspace*{-2.0cm}
\scalebox{0.4}{\includegraphics{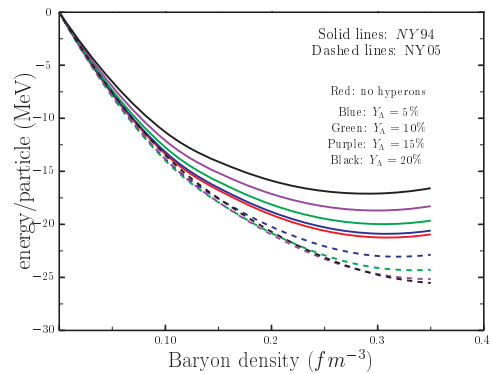}}
\vspace*{-2.0cm}
\caption{(Color online) Energy/particle as a function of density in symmetric nuclear matter for various
$\Lambda$ concentrations. Predictions obtained from BHF calculations. The solid(dashed) lines
are obtained with the NY94(NY05) potentials. 
} 
\label{NY3}
\end{center}
\end{figure}

This is further confirmed in 
Fig.~\ref{NY3}, where we show predictions from conventional BHF calculations (i.e., no ``Dirac" effects). 
As in the previous figures, 
the red curve is the nucleonic EoS. The solid(dashed) curves are the predictions with NY94(NY05).  
Clearly, the effects are opposite depending on the NY interaction. Qualitatively, the trend seen
in each group of curves (solid or dash) is approximately consistent with the one observed previously in the
corresponding (i.~e. same NY potential) DBHF predictions. However, the ``cross over" of the curves                       
seen in Fig.~\ref{NY1} at about twice saturation density is due to a large extent to the more repulsive nature of the
nucleonic EoS in the DBHF calculation (see comments above). 

In conclusion, the effect of hyperons on the EoS is strongly dependent upon the baseline
(nucleonic) EoS as well as the NY potential model.                                              
With regard to the first issue, we stress the importance of starting from a realistic EoS, such as the one
predicted in DBHF calculations or obtained with the inclusion of three-body forces. 
The second observation confirms the conclusions of Ref.\cite{vlowk}                
concerning the large uncertainties originating from the bare NY potentials.
Unfortunately, the existing data do not set sufficient constrains on the potentials, as demonstrated by
the fact that different potentials can fit the available scattering data equally accurately but
produce very different scattering lengths \cite{vlowk}.


Of course,        
the actual fraction of hyperons present in star matter must be determined by the equations
of $\beta$ stability and charge neutrality for highly asymmetric matter containing neutrons, protons, hyperons,
and leptons as described in Sec.~5.1.                  
What we have learnt at this time is 
that the predicted energy/particle in symmetric matter is very sensitive to the 
chosen NY interaction.                                                                   
The uncertainties due to the model dependence discussed in this 
paper are likely to impact any conclusions on the properties of strange neutron stars,
which therefore must be interpreted with caution. These include considerations of deconfinement
and possible transition from hadronic to quark matter, which depend sensitively on the equation 
of state in the hadronic phase. 
\begin{figure}[!t] 
\centering          
\vspace*{-3.2cm}
\includegraphics[totalheight=5.5in]{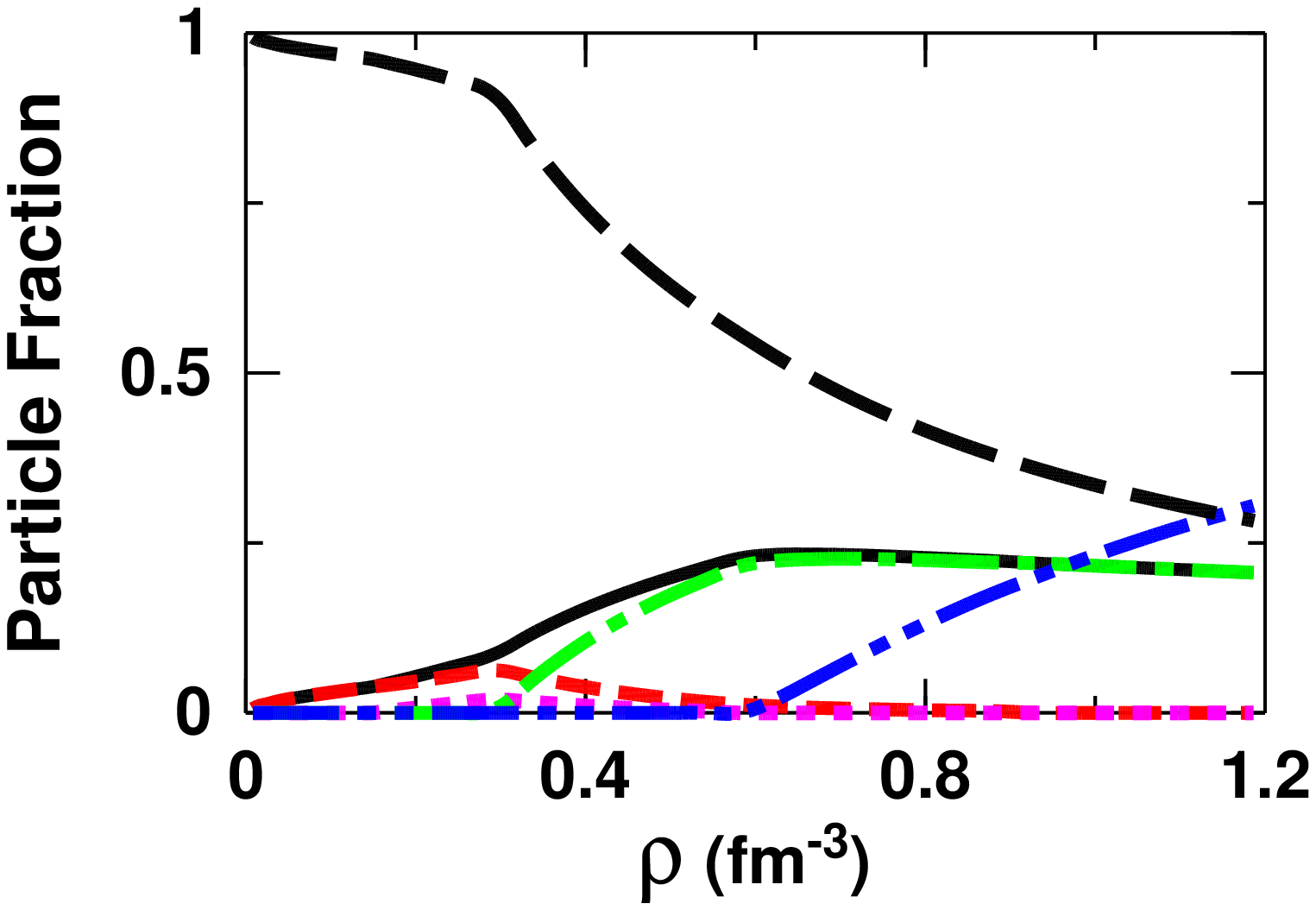}                
\vspace*{-3.5cm}
\caption{(color online) Various particle fractions in $\beta$-stable matter.                        
Neutron: long-dashed black; proton: solid black; $\Sigma ^-$: dash-dotted green; 
$\Lambda$: dash-double-dotted blue; electrons: short-dashed red; muons: dotted purple. 
} 
\label{allfrac}
\end{figure}

\subsection{Neutron star matter with non-interacting hyperons}
As anticipated at the end of Sec.~5.2,                      
the uncertainties that we have demonstrated in Sec.~5.3                               
suggest that       
present NY interactions are not sufficiently constrained to allow for robust conclusions 
concerning the occurrence of a hybrid phase. Therefore, to avoid those uncertainties, 
 we will now try to {\it estimate} the impact of hyperons on the $\beta$-stable 
DBHF EoS 
considering {\it non-interacting} hyperons.                                               
Equations~(\ref{nybeta1},\ref{nybeta1b},\ref{nybeta2}) 
are solved (for fixed total baryon density) treating $\Lambda$ and $\Sigma ^-$ as free
(non-relativistic) fermions. The nucleonic energies are taken from our DBHF EoS for IANM, evaluated at 
some nucleon density, $\rho_n + \rho_p$, with                                                         
$\rho_n$ and $ \rho_p$ to be determined along with the other particle fractions. (To allow for the solution of 
the 6$\times$6 (including the normalization condition) system of algebraic equations for the six unknown
particle concentrations, our nucleonic EoS is first parametrized in terms of analytic functions of density.)
Those fractions are shown in Fig.~\ref{allfrac}. The trend is very similar to the one shown in Ref.\cite{nycatania3}
with free hyperons, which, in turn, is qualitatively consistent with the one seen when the NY interaction
is turned on.\cite{nycatania3} The onset of the $\Sigma^-$ baryon occurs at a density of about 0.3$fm^{-3}$. Leptons start
to disappear from the system after hyperon formation, an indication that charge neutrality through 
$\Sigma^-$ production is energetically more favorable than through $\beta$ decay.       
The onset of $\Lambda$ production occurs at a density of approximately 0.6$fm^{-3}$. These thresholds are in 
good agreement with those reported in Ref.\cite{nycatania3} for both free and interacting hyperons, when 
the BHF+TBF predictions are considered (the TBF shifts down the threshold densities for $\Lambda$ and 
$\Sigma^-$ formation as compared with the case of two-body forces only). 

\begin{figure}[!t] 
\centering          
\includegraphics[totalheight=4.0in]{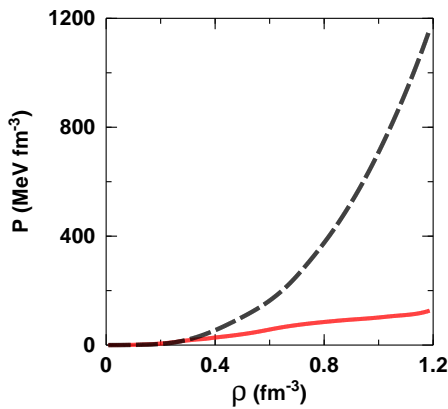}                
\vspace*{-2.5cm}
\caption{(color online)  Pressure in $\beta$-stable matter from the DBHF calculation                
 with  no hyperons (black dash) and 
 with free hyperons (solid red).           
} 
\label{Ypr}
\end{figure}

In Fig.~\ref{Ypr}, we show the pressure in $\beta$-stable matter with the Idaho DBHF EoS with and 
without (free) hyperons. The differences are dramatic but consistent with previous observations.\cite{nycatania3}
The effect of including the NY interaction is quite small compared with the differences seen in 
 Fig.~\ref{Ypr}. Thus, one conclusion seems appropriate at this point: Regardless the chosen NY interaction,
the effect of adding strange baryons is a dramatic softening of the EoS, (consistent with a large reduction of the 
neutron star maximum mass). {\it Furthermore, given the uncertainties arising from the NY potential 
demonstrated in the previous section, the effect of free hyperons may very well provide a realistic, average estimate
of the impact of including strangeness.} 

\begin{figure}[!t] 
\centering          
\includegraphics[totalheight=4.0in]{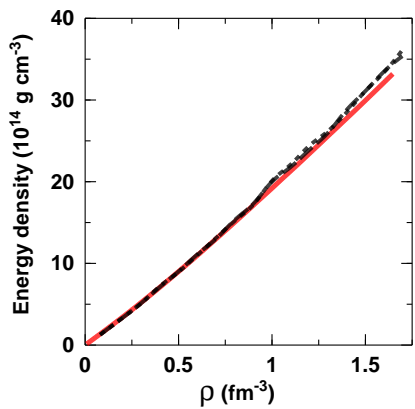}                
\vspace*{-2.5cm}
\caption{(color online)  Energy density in $\beta$-stable matter from various models: The black dashed and 
dotted lines (essentially identical) are the results of BHF calculations with the Argonne V18 and the Paris potentials,
respectively, including TBF and hyperons interacting via the Nijmegen soft-core NY potential (see 
Ref.$^{132}$). The red solid line is obtained with the DBHF calculation for the nucleonic EoS and
free hyperons.} 
\label{Yedens}
\end{figure}

In Fig.~\ref{Yedens} we compare energy densities in $\beta$-stable matter as obtained from the Idaho
DBHF EoS with free hyperons and from the two models (AV18 + TBF, Paris + TBF) used in Ref.\cite{nycatania3}
along with the NSC NY potential.\cite{Nij89} The similarity between the curves indicates that                        
the effect of including the NY interaction in star matter is relatively small and/or that differences in the respective 
nucleonic EoS may compensate those which originate from the treatment of hyperons. 

The neutron star maximum mass we obtain with the DBHF EoS and free hyperons is close to 1.2 $M_{\odot}$.
Considering that inclusion of the NY interaction would increase the pressure at high density, this result is
not inconsistent with those shown in Table 6. 
In summary, our analysis confirms a large softening of the $\beta$-stable EoS as a consequence of including strange
baryons, accompanied by a strong reduction of the maximum mass, a conclusion that appears to be nearly model independent.  
The softening is 
mostly caused by  conversion of the kinetic energy of the existing particle species into masses of the newly 
produced particles. 
The resulting maximum mass reduction 
appears in contraddiction with present constraints, and thus 
leaves intriguing questions open concerning phase transitions and degrees 
of freedom appropriate for high-density stellar matter.

\section{Conclusions and Outlook}

The EoS of hadronic matter enters in a variety of systems, from atomic nuclei to the most compact
form of matter found in the universe, namely the matter in the interior of neutron stars. 
EoS-related properties can and should be calculated {\it ab initio} and consistently from the same 
basic nuclear forces. They include: the nuclear matter ``optical potential", effective in-medium cross sections,
nucleon mean free path in nuclear matter, neutron skins, and neutron star properties.               

We have reviewed recent calculations of the EoS, 
over a large range of densities and isospin asymmetries, 
with a particular eye on the microscopic approach.                                   
The obvious advantage of the latter lies in the opportunity of interpreting the predictions in terms
of the input nuclear forces and their behavior in the many-body environment.
In turn, this facilitates the identification of potentially missing dynamics.        

We have discussed differences among microscopic models which do or do not include explicit
TBF. 
It is quite clear that model dependence amongst predictions can be quite large. 
Naturally, this is especially the case at those densities where constraints are the weakest.

Rich and diverse effort is presently going on 
to improve the available constraints on the EoS or find new ones.                   
These constraints are usually extracted through the analysis of selected heavy-ion collision
observables. At the same time,   
partnership between nuclear physics and astrophysics is becoming increasingly important towards 
advancing our understanding of exotic matter. 
Perhaps the best prospects to discriminate among families of EoS, especially at high density, 
are in more accurate measurements of neutron stars radii and/or
moments of inertia. 

Other EoS-related issues of contemporary interest which we have not reviewed here include 
temperature dependence and polarizability of nuclear/neutron matter. 
Knowledge of the finite-temperature EoS plays a crucial role in the final stages of a supernova evolution.
Microscopic models which can successfully describe the ground state of nuclear matter 
should be the starting point to move on to its excited states. 

As a final note, 
the emergence of FRIB\cite{FRIB} will mark a turning point in the progress of experimental nuclear physics, 
possibly unveiling unknown 
areas on the nuclear chart. 
It is thus imperative that such large-scale projects
be constantly supported by theoretical calculations with predictive power.

\section*{Acknowledgements}
Support from the U.S. Department of Energy under grant DE-FG02-03ER41270  
is gratefully acknowledged. 
I like to thank D. Alonso and P. Liu for their help in various phases of the 
Idaho project. 
I am grateful to M. Baldo, U. Lombardo, and 
H.-J. Schulze for useful communications, to F. Weber for providing the 
crustal EoS, and to J. Heidenbauer for providing the nucleon-hyperon potential
code.

\end{document}